\documentclass[prd,twocolumn,showpacs,floatfix,amsmath,nofootinbib,amssymb,floatfix]{revtex4}
\usepackage{graphicx,color,dcolumn,booktabs,bm,multirow}
\usepackage{longtable,lscape}
\usepackage{txfonts}
\usepackage{enumitem}
\usepackage{overpic}
\usepackage{amssymb}
\usepackage{indentfirst}
\usepackage{feynmf}   %{feynmp}
\usepackage{slashed}  %for Feynman symbols
\usepackage{cases}
\usepackage{color}
\usepackage{multirow}
\usepackage{epstopdf}
\usepackage{graphicx,color,dcolumn,booktabs,bm}
\usepackage[colorlinks,
citecolor=blue,
anchorcolor=red,
menucolor=red,
linkcolor=red,
filecolor=red,
runcolor=red,
urlcolor=blue,
frenchlinks=red]{hyperref}
\usepackage{amsmath}
\usepackage{mathrsfs}
\usepackage{epsfig}
\usepackage{youngtab}
\usepackage{graphicx}
\usepackage{txfonts}
\usepackage{cleveref}

\begin{document}
\title{A study on the properties of hidden-charm pentaquarks with double strangeness}

\author{Xuejie Liu$^1$}\email[E-mail: ]{1830592517@qq.com}
\author{Yue Tan$^{2}$}\email[E-mail:]{tanyue@ycit.edu.cn}
\author{Xiaoyun Chen$^{5}$}\email[E-mail:]{xychen@jit.edu.cn}
\author{Dianyong Chen$^{3,6}$\footnote{Corresponding author}}\email[E-mail:]{chendy@seu.edu.cn}
\author{Hongxia Huang$^4$}\email[E-mail:]{hxhuang@njnu.edu.cn}
\author{Jialun Ping$^4$}\email[E-mail: ]{jlping@njnu.edu.cn}
\affiliation{$^1$School of Physics, Henan Normal University, Xinxiang 453007, P. R. China}
\affiliation{$^2$School of Mathematics and Physics, Yancheng Institute of Technology, Yancheng, 224051,  P. R. China}
\affiliation{$^3$Lanzhou Center for Theoretical Physics, Lanzhou University, Lanzhou 730000, P. R. China}
\affiliation{$^4$Department of Physics, Nanjing Normal University, Nanjing 210023, P. R. China}
\affiliation{$^5$College of Science, Jinling Institute of Technology, Nanjing 211169, P. R. China}
\affiliation{$^6$School of Physics, Southeast University, Nanjing 210094, P. R. China}

\begin{abstract}
Since the LHCb collaboration successively reported the hidden-charm exotic states $P_{c}$ and $P_{cs}$, these exotic states have rightfully attracted significant attraction from theoretical perspectives. Motivated by these exotic states, this study systematically investigates the hidden-charm double-strange pentaquark system with the quark configuration $nssc\bar{c}$ using the resonating group method within the framework of the quark delocalization color screening model.  The dynamical estimations, incorporating channel coupling, identify five resonance states in the hidden-charm double-strange systems, denoted as $\Xi_{c}^{\prime} \bar{D}_{s}^{\ast}$ and $\Xi^{\ast}_{c}\bar{D}_{s}^{\ast}$ with $J^{P}=1/2^{-}$,  $\Xi^{\ast}_{c}\bar{D}_{s}$, $\Xi^{\ast}_{c}\bar{D}_{s}^{\ast}$, and $\Omega^{\ast}\bar{D}^{\ast}$ with  $J^{P}=3/2^{-}$, through scattering phase shifts processes of various channels. $\Xi_{c}^{\prime} \bar{D}_{s}^{\ast}$ and $\Xi^{\ast}_{c}\bar{D}_{s}^{\ast}$ with $J^{P}=1/2^{-}$, with mass range and decay widths of (4684 MeV$-$4688 MeV, 6.4 MeV$-$24.2 MeV) and (4751 MeV$-$4756 MeV, 3.2 MeV$-$18.7 MeV), can be derived from the scattering processes of channels $\Xi \eta_{c}$, $\Xi J/\psi$ and $\Xi_{c}\bar{D}_{s}$, it should be noted that resonance state $\Xi^{\prime}_{c}\bar{D}_{s}^{\ast}$, in addition to the channels mentioned above, can also be observed in $\Xi^{\ast}J/\psi$ and $\Xi_{c}\bar{D}^{\ast}_{s}$. For the quantum number $J^{P}=3/2^{-}$,  $\Xi^{\ast}_{c}\bar{D}_{s}^{\ast}$ and $\Omega^{\ast}\bar{D}^{\ast}$ can be identified in channels $\Xi J/\psi$, $\Xi^{\ast} J/\psi$ and $\Xi_{c}\bar{D}_{s}^{\ast}$, and $\Xi^{\ast} J/\psi$ and $\Xi_{c}^{\prime}\bar{D}_{s}^{\ast}$, respectively,  the estimated masses and decay widths of these resonance states range from approximately (4749 MeV$-$4755 MeV, 16.4 MeV) and (4771 MeV-4772 MeV, 18.8 MeV).  However,  the residual resonance state $\Xi^{\ast}_{c}\bar{D}_{s}$, for which the estimated mass and decay width are about 4600 MeV and 21.7 MeV, is expected to be found only in channel $\Xi_{c}\bar{D}_{s}^{\ast}$. Given the current theoretical estimations, experiments are anticipated to further explore hidden-charm double-strange pentaquark states through the possible decay channels, thereby offering a strong theoretical foundation for their experimental investigation.

\end{abstract}

\maketitle

\section{\label{sec:introduction}Introduction}
Over the past few decades, with continuous advancements in experimental equipment and techniques, a growing number of exotic hadron states, which differ from conventional mesons made of quark-antiquark pairs and baryons composed of three quarks, have been detected. Indeed,  there has been ongoing discourse on non-conventional exotic states since the birth of the quark model~\cite{Gell-Mann:1964ewy}. Although their existence and properties have been investigated thoroughly since then from different points of view using different approaches in both experiments and theory, the definitive observational evidence first emerged in 2003 when the Belle Collaboration observed the tetraquark state X(3872)~\cite{Belle:2003nnu}, later confirmed by other collaborations, such as the BaBAR Collaboration~\cite{BaBar:2004oro}, CDF II Collaboration ~\cite{CDF:2003cab, CDF:2009nxk}, D0 Collaboration ~\cite{D0:2004zmu}, LHCb Collaboration~\cite{LHCb:2011zzp}, and CMS Collaboration ~\cite{CMS:2013fpt}. With the discovered exotic state X(3872) as dividing line, a series of exotic states are successively reported by various experimental groups with more experimental data with high precision as summarized in reviews~\cite{Liu:2013waa, Hosaka:2016pey, Chen:2016qju, Richard:2016eis, Lebed:2016hpi, Olsen:2017bmm, Guo:2017jvc, Liu:2019zoy, Brambilla:2019esw, Meng:2022ozq, Chen:2022asf}.  The discovery of exotic states has sparked extensive research into their properties and internal structure. In the broader perspective, exploring exotic states not only sheds new light on the exotic states' structure but also provides valuable hints for further understanding the non-perturbative behavior of the quantum chromodynamics (QCD) in the low-energy region.

Among the various exotic states discovered, the hidden-charm pentaquark states have attracted the most attention. In 2015,  two unique exotic states, $P_{c}(4380)$ and $P_{c}(4450)$, named pentaquark states containing five valence quarks, were announced by the LHCb Collaboration in the $J/\psi p$ mass distribution of the decay of $\Lambda_{b}\rightarrow  J/\psi p K$ ~\cite{LHCb:2015yax}, this two $P_{c}$ states were also analyzed in the full amplitude fit to the decay of $\Lambda_{b}\rightarrow J/\psi p \pi$ with a low significance~\cite{LHCb:2016lve}. In 2019,  the LHCb Collaboration identified three new exotic states, referred to as $P_{c}(4312)$, $P_{c}(4440)$ and $P_{c}(4457)$, in the same process by analyzing a larger data sample, with $P_{c}(4440)$ and $P_{c}(4457)$ resultsing from the splitting of $P_{c}(4450)$~\cite{LHCb:2019kea}.  Subsequently,  the hidden-charm pentaquark state $P_{c}(4337)$ was observed in the $J/ \Psi p$ and $J/ \psi \bar{p}$ mass distribution of the decay $B_{s}\rightarrow J/\psi p \bar{p}$~\cite{LHCb:2021chn}.  Additionally, the LHCb Collaboration also performed detection in process of $\Lambda_{b}\rightarrow \eta_{c} p K$ for the investigation of $P_{c}(4312)$, however, no signal of $P_{c}(4312)$ was observed in the $\eta_{c} p$ mass distribution~\cite{LHCb:2020kkc}.  In 2020,  the hidden-charm strange exotic state  $P_{cs}(4459)$, which can also be viewed as a mixture of two pentaquark states near the mass of $P_{cs}(4459)$, was found in the $J/ \psi \Lambda$ mass distribution in the decay of $\Xi_{b}\rightarrow J/ \psi \Lambda K$~\cite{LHCb:2020jpq}.  Very recently, the LHCb Collaboration observed a new structure $P_{cs}(4338)$ in the $J/ \psi \Lambda$ mass distribution in the $B^{-}\rightarrow J/ \psi \Lambda p$ decay~\cite{LHCb:2022ogu}.

For these hidden-charm exotic states observed,  there are various possibilities about their properties and structures due to the uncertainties of these exotic states' nature.  Theoretically,  there has been a plethora of studies trying to unveil the nature of these resonance states with hidden-charm. Among the most commonly used possible substructures is the molecular interpretation since the observed $P_{c}$ and $P_{cs}$ are below the threshold of the corresponding di-hadron systems from several to several tens MeVs. Based on the molecular interpretation, many theoretical models,  the one-pion-exchange model~\cite{Chen:2016heh}, the sum rules method~\cite{Chen:2016otp, Azizi:2016dhy, Azizi:2018bdv, Azizi:2020ogm}, the quaipotential Bethe-Salpeter equation approach~\cite{He:2019ify, Zhu:2021lhd}, the contact-range effective field theory approach~\cite{Liu:2019tjn}, the effective Lagrangian approach~\cite{Xiao:2019mvs, Lu:2016nnt}, the constituent chiral quark model~\cite{Hu:2021nvs}, and the chiral perturbation theory~\cite{Meng:2019dba}, were put forward.  The meson baryon molecular interpretation was also adopted in Refs.~\cite{Chen:2015loa, Chen:2015moa, He:2015cea, Meissner:2015mza, Roca:2015dva, Chen:2020opr, Yan:2021nio, Wu:2021caw,Chen:2020uif, Phumphan:2021tta,Du:2021fmf, Lu:2021irg,Gao:2021hmv, Yalikun:2021dpk}. In addition to the molecular interpretations, there exist other explanations for these pentaquark states,  such as hadron-charmonia~\cite{Eides:2019tgv}, compact pentaquark states~\cite{Ali:2019npk, Wang:2019got, Cheng:2019obk, Weng:2019ynv, Zhu:2019iwm, Pimikov:2019dyr, Ruangyoo:2021aoi}, virtual states~\cite{Fernandez-Ramirez:2019koa}, triangle singularities~\cite{Nakamura:2021qvy}, and cusp effects~\cite{Burns:2022uiv}. 

%%%%%%%%%%%%%%%%%%%%%%%%%%%%%%%%%%%%%%%%%%%%%%%%%%%%%%%%%
\begin{figure}[t]
    \includegraphics[scale=0.35]{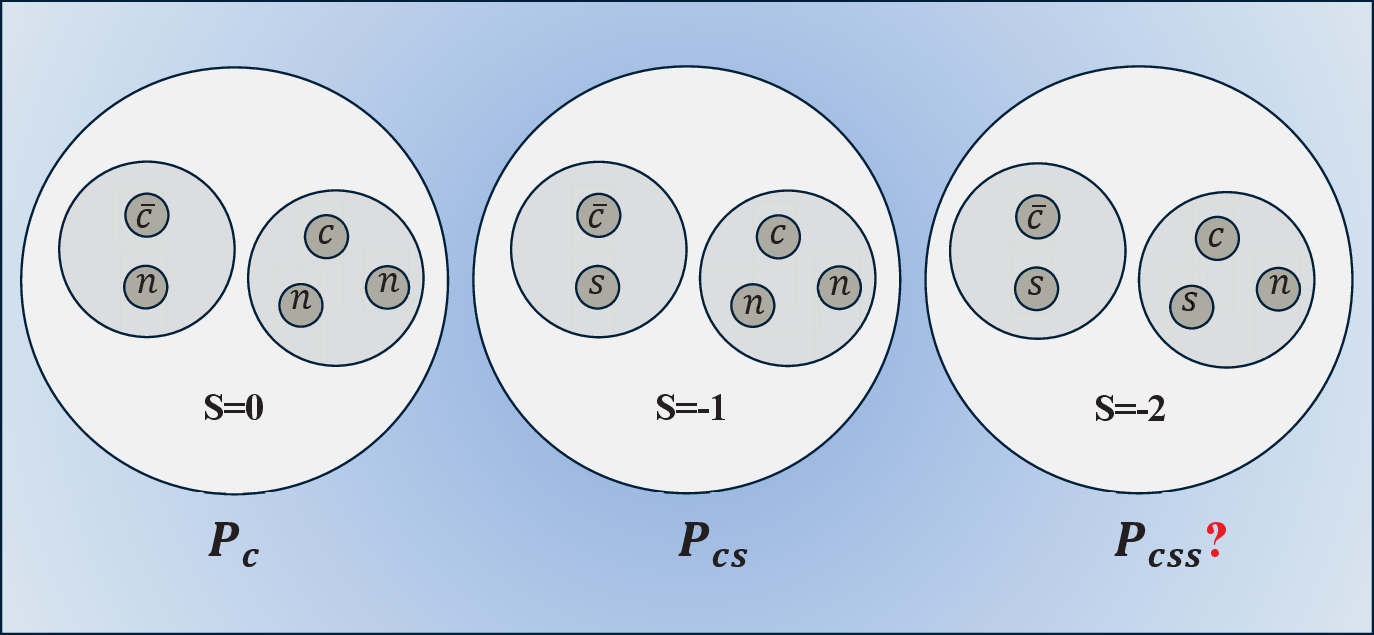}%{hidden-charm-family.eps}
   % \vspace{5cm}
    \caption{ The hidden-charm pentaquark family with considering the different strangeness number, "n" represents the light the quark u or d.}
    \label{Fig1}
\end{figure}
%%%%%%%%%%%%%%%%%%%%%%%%%%%%%%%%%%%%%%%%%%%%%%%%%%%%%%%%%

Given that a series of hidden-charm pentaquark states $P_{c}$ and $P_{cs}$  with varying strangeness numbers were detected by the LHCb Collaboration,  as illustrated in Fig.~\ref{Fig1},  a logical step would be to explore whether there exists a hidden-charm double-strange pentaquark state within the hidden-charm pentaquark family.  The CMS Collaboration recently investigated the $J/ \psi \Xi$ invariant mass spectrum from  $\Lambda_{b}^{0}\rightarrow J/ \psi \Xi^{-} K^{+}$~\cite{CMS:2024vnm}, this decay channel is of particular interest as it can proceed through exotic intermediate resonances, such as pentaquark states. However, low yield and poor resolution precluded observation of a clear spectrum in the $J/ \psi \Xi$ invariant mass spectrum.  Even so, this observation is significant for the study of double-strange pentaquarks, providing new avenues for exploring the properties and interactions of these exotic states. In Refs.~\cite{Wang:2020bjt, Marse-Valera:2022khy, Roca:2024nsi, Marse-Valera:2024apc, Magas:2024biu, Ortega:2022uyu}, several hidden-charm double-strange pentaquark states from the molecular perspective were obtained. In Ref.~\cite{Wang:2020bjt}, a pair of slightly bound meson-baryon molecules with double strangeness, $\Xi^{\prime}_{c}\bar{D}_{s}^{\ast}$ with $J^{P}=3/2^{-}$ and $\Xi^{\ast}_{c}\bar{D}_{s}^{\ast}$ with $J^{P}=5/2^{-}$,  were derived in the One-boson-exchange model. In Refs.~\cite{Marse-Valera:2022khy, Marse-Valera:2024apc,Magas:2024biu},  a pole, $P_{css}(4493)$ with $J^{P}=1/2^{-}$, coupled most strongly to the $\Omega_{c}\bar{D}$ and $\Xi^{\prime}_{c}\bar{D}_{s}$ channels. The other one, $P_{css}(4633)$ with $J^{P}=1/2^{-}$ or $J^{P}=3/2^{-}$ was obtained from the interaction of vector mesons with baryons, and coupled dominantly to the $\Omega_{c}\bar{D}^{\ast}$ and $\Xi^{\prime}_{c}\bar{D}_{s}$ channels.  Similar results can also be found in Ref.~\cite{Roca:2024nsi}, and two more states of  $PB^{\ast}$ and $VB^{\ast}$ nature were also found. Besides, eight candidates were predicted as $\Xi_{c}^{(\prime,\ast)}\bar{D}_{s}^{(\prime,\ast)}$ in different quantum number with $I=1/2$ in a constituent quark model~\cite{Ortega:2022uyu}.  The diquark model~\cite{Shi:2021wyt, Ozdem:2024usw}, QCD sum rule~\cite{Azizi:2021pbh} and the hadron-quarkonium model~\cite{Ferretti:2020ewe, Ferretti:2021zis} were also used to search for the hidden-charm double-strange pentaquarks.  In addition to the above motioned the recent work of Ref.~\cite{Oset:2024fbk} looked for the $P_{css}$ with $I(J^{P})=1/2(1/2^{-})$ generated via the pseudoscalar-baryon interaction through the decay of $\Xi_{b}^{0}\rightarrow \eta \eta_{c}\Xi^{0}$ and $\Omega_{b}^{-}\rightarrow K^{-} \eta_{c} \Xi^{0}$.  The $\Xi_{b}\rightarrow \Xi J/\psi \phi$ decay process was estimated and the signature of the $S=-2$ pentaquarks were expected to be seen in the $J /\psi \Xi$ mass distribution~\cite{Marse-Valera:2024apc}. However, no theoretical evidence of resonance or bound states with hidden-charm double-strange was found in ~\cite{Wu:2010vk, Xiao:2013yca} using the coupled channels $\Xi_{c}^{\prime} \bar{D}_{s}$,  $\Xi_{c} \bar{D}_{s}$ and $\Omega_{c} \bar{D}$ likely due to the neglect of some important nondiagonal channels.

In the present work, the nature of the possible hidden-charm double-strange pentaquark states with quark content $nssc\bar{c}$ are also investigated in the framework of the quark delocalization color screening model.  The possible physical channels with different quantum numbers are considered, and then the effective potential of each channel has been systematically estimated. Subsequently, to confirm whether the physical channel with an effective attractive potential yields a bound-state solution through the resonance group method in the dynamical estimations, the corresponding single-channel and channel-coupling estimations are applied to this system. Based on these analyses, and by incorporating the constraints of quantum number conservation and phase shifts space, we explore possible resonance states in possible hidden-charm double-strange scattering decay processes. 

This work is organized as follows. In the next section, the detail of the QDCSM is presented.  In section~\ref{dis}, a comprehensive numerical analysis of the hidden-charm pentaquark with double strangeness is carried out, including the effective potential, possible bound states, and resonance states. Finally, section~\ref{sum} summarizes the main conclusions of this present work.

\section{THE QUARK DELOCALIZATION COLOR SCREENING MODEL  \label{mod}}
%In this work, the main purpose is to detect the presence of possible bound states or resonance states in the charmed-strange pentaquark state system. Now, we use the quark delocalization color screening model to calculate the spectra of the the $N\bar{D}$ system. Besides, we employ the resonating group (RGM) method to calculate the baryon-meson scattering phase shifts and to look for the resonance states. %In the following, we provide a detailed description of the model and method.

The QDCSM is an extension of the native quark cluster model~\cite{DeRujula:1975qlm, Isgur:1979be, Isgur:1978wd, Isgur:1978xj} and was developed to address multiquark systems (More detail of QDCSM can be found in the Refs.~\cite{Wang:1992wi, Chen:2007qn, Chen:2011zzb, Wu:1996fm, Huang:2011kf}).
In the QDCSM, the general form of the Hamiltonian for the pentaquark system is,
\begin{equation}
H = \sum_{i=1}^{5} \left(m_i+\frac{\boldsymbol{p}_i^2}{2m_i}\right)-T_{\mathrm{CM}}+\sum_{j>i=1}^5V(r_{ij}),\\
\end{equation}
where the center-of-mass kinetic energy, $T_{\mathrm{CM}}$, is subtracted without losing generality since we mainly focus on the internal relative motions of the multiquark system. The two body potentials include the color-confining potential, $V_{\mathrm{CON}}$, one-gluon exchange potential, $V_{\mathrm{OGE}}$, and Goldstone-boson exchange potential, $V_{\chi}$, respectively, i.e.,
\begin{equation}
V(r_{ij}) = V_{\mathrm{CON}}(r_{ij})+V_{\mathrm{OGE}}(r_{ij})+V_{\chi}(r_{ij}).
\end{equation}

Noted herein that the potentials include the central, spin-spin, spin-orbit, and tensor contributions, respectively. Since the current calculation is based on the S$-$wave, only the first two kinds of potentials will be considered attending the goal of the present calculation and for clarity in our discussion. In particular, the one-gluon-exchange potential, $V_{\mathrm{OGE}}(r_{ij})$, reads,
\begin{eqnarray}
V_{\mathrm{OGE}}(r_{ij}) &=& \frac{1}{4}\alpha_{s} \boldsymbol{\lambda}^{c}_i \cdot\boldsymbol{\lambda}^{c}_j \nonumber\\
&&\times\left[\frac{1}{r_{ij}}-\frac{\pi}{2}\delta(\boldsymbol{r}_{ij})\left(\frac{1}{m^2_i}+\frac{1}{m^2_j}
+\frac{4\boldsymbol{\sigma}_i\cdot\boldsymbol{\sigma}_j}{3m_im_j}\right)\right],\ \
\end{eqnarray}
where $m_{i}$ is the quark mass, $\boldsymbol{\sigma}$ and $\boldsymbol{\lambda^{c}}$ are the Pauli matrices and SU(3) color matrices, respectively. The QCD-inspired effective scale-dependent strong coupling constant, $\alpha_{s}$, offers a consistent description of mesons and baryons from the light to the heavy quark sectors, which can be written by,
\begin{equation}
\alpha_{s}(\mu)= \frac{\alpha_{0}}{\ln(\frac{\mu^{2}+\mu_{0}^{2}}{\Lambda_{0}^2})}.
\end{equation}

In the QDCSM, the confining interaction $V_{\mathrm{CON}}(r_{ij})$ can be expressed as
\begin{equation}
 V_{\mathrm{CON}}(r_{ij}) =  -a_{c}\boldsymbol{\lambda^{c}_{i}\cdot\lambda^{c}_{j}}\Big[f(r_{ij})+V_{0_{ij}}\Big] \ ,
\end{equation}
where $a_{c}$ represents the strength of the confinement potential and $V_{0_{ij}}$ refers to the zero-point potential. Moreover, in the quark delocalization color screening model, the quarks in the considered pentaquark state involving quark components of $nssc\bar{c}$ are first divided into two clusters, which are baryon cluster composed of three quarks, and meson cluster composed of one quark and one antiquark. And then the five-body problem can be simplified as a two-body problem the $f(r_{ij})$ is,
\begin{equation}
 f(r_{ij}) =  \left\{ \begin{array}{ll}r_{ij}^2 & \quad \mbox{if }i,j\mbox{ occur in the same cluster}, \\
\frac{1 - e^{-\mu_{ij} r_{ij}^2} }{\mu_{ij}} & \quad \mbox{if }i,j\mbox{ occur in different cluster},
\end{array} \right.
\label{Eq:fr}
\end{equation}
where the color screening parameter $\mu_{ij}$ is determined by fitting the deuteron properties, nucleon-nucleon, and nucleon-hyperon scattering phase shifts~\cite{Chen:2011zzb, Wang:1998nk}, with $\mu_{nn}= 0.45\ \mathrm{fm}^{-2}$, $\mu_{ns}= 0.19\ \mathrm{fm}^{-2}$
and $\mu_{ss}= 0.08\ \mathrm{fm}^{-2}$, satisfying the relation $\mu_{ns}^{2}=\mu_{nn}\mu_{ss}$, where $n$ represents $u$ or $d$ quark. From this relation, a fact can be found that the heavier the quark, the smaller the parameter $\mu_{ij}$. When extending to the heavy-quark case, there is no experimental data available, so we take it as an adjustable parameter. In Ref.~\cite{Huang:2015uda}, we investigate the mass spectrum of $P_{\psi}^N$ with $\mu_{cc}$ varying from $10^{-4}$ to $10^{-2}$ fm$^{-2}$, and our estimation indicated that the dependence of the parameter $\mu_{cc}$ is not very significant. In the present work, we take $\mu_{cc}=0.01\ \mathrm{fm}^{-2}$. Then $\mu_{sc}$ and $\mu_{nc}$ are obtained by the relations $\mu_{sc}^{2}=\mu_{ss}\mu_{cc} $ and $\mu_{nc}^{2}=\mu_{nn}\mu_{cc}$, respectively. It should be noted that $\mu_{ij}$ are phenomenal model parameters, their values are determined by reproducing the relevant mass spectra and phase shifts of the scattering processes. In Ref.~\cite{Chen:2011zzb}, the authors found that with the relation $\mu_{qs}^2=\mu_{qq} \mu_{ss}$, the masses of the ground state baryons composed of light quarks could be well reproduced. Later on, such relations have been successfully applied to investigate the states with heavy quarks~\cite{Huang:2013rla, Huang:2015uda, Liu:2022vyy}.

% Besides, as indicated in Ref.~\cite{Huang:2011kf}, the phenomenological color screening confinement is an effective description of the hidden color channel coupling, so the hidden color channels of the pentaquark system in QDCSM are excluded.

The Goldstone-boson exchange interactions between light quarks appear because of the dynamical breaking of chiral symmetry. The following $\pi$, $K$, and $\eta$ exchange terms work between the chiral quark-(anti)quark pair, which read,
\begin{eqnarray}
V_{\chi}(r_{ij}) & =&  v^{\pi}_{ij}(r_{ij})\sum_{a=1}^{3}\lambda_{i}^{a}\lambda_{j}^{a}+v^{K}_{ij}(r_{ij})\sum_{a=4}^{7}\lambda_{i}^{a}\lambda_{j}^{a}+v^{\eta}_{ij}(r_{ij})\nonumber\\
&&\left[\left(\lambda _{i}^{8}\cdot
\lambda _{j}^{8}\right)\cos\theta_P-\left(\lambda _{i}^{0}\cdot
\lambda_{j}^{0}\right) \sin\theta_P\right], \label{sala-Vchi1}
\end{eqnarray}
with
\begin{eqnarray}
  v^{B}_{ij} &=&  {\frac{g_{ch}^{2}}{{4\pi}}}{\frac{m_{B}^{2}}{{\
12m_{i}m_{j}}}}{\frac{\Lambda _{B}^{2}}{{\Lambda _{B}^{2}-m_{B}^{2}}}}
m_{B}     \nonumber    \\
&&\times\left\{(\boldsymbol{\sigma}_{i}\cdot\boldsymbol{\sigma}_{j})
\left[ Y(m_{B}\,r_{ij})-{\frac{\Lambda_{B}^{3}}{m_{B}^{3}}}
Y(\Lambda _{B}\,r_{ij})\right] \right\},
\end{eqnarray}
with $B=(\pi, K,  \eta)$ and $Y(x)=e^{-x}/x$ to be the standard Yukawa function. $\boldsymbol{\lambda^{a}}$ is the SU(3) flavor Gell-Mann matrix. The masses of the $\eta$, $K$, and $\pi$ meson are taken from the experimental value~\cite{ParticleDataGroup:2018ovx}. By matching the pion exchange diagram of the $NN$ elastic scattering process at the quark level and the hadron level, one can relate the $\pi qq$ coupling with the one of $\pi NN$, which is~\cite{Vijande:2004he, Fernandez:1986zn},
\begin{equation}
\frac{g_{ch}^{2}}{4\pi}=\left(\frac{3}{5}\right)^{2} \frac{g_{\pi NN}^{2}}{4\pi} {\frac{m_{u,d}^{2}}{m_{N}^{2}}},
\end{equation}
which assumes that the flavor SU(3) is an exact symmetry, and only broken by the masses of the strange quark. As for the coupling $g_{\pi NN}$, it was determined by the $NN$ elastic scattering~\cite{Fernandez:1986zn}. All model parameters are taken from Ref.~\cite{Huang:2015uda}, which were determined by reproducing the mass spectrum of the ground states mesons and baryons in QDCSM.  According to the model parameters obtained from Refs.~\cite{Huang:2015uda}, in which the hidden-charm pentaquark states $P_{c}$ can be well explained as molecular states,  so the properties of hidden-charm double-strange pentaquark systems will be examined in the next section.

Besides, in QDCSM, the quark delocalization is realized by specifying the single-particle orbital wave function as a linear combination of left and right Gaussian basis, the single--
particle orbital wave functions used in the ordinary quark cluster model reads,
\begin{eqnarray}\label{wave0}
\psi_{\alpha}(\boldsymbol{s}_{i},\epsilon)&=&\left(\Phi_{\alpha}(\boldsymbol{s}_{i})
  +\epsilon\Phi_{\beta}(\boldsymbol{s}_{i})\right)/N(\epsilon), \nonumber \\
\psi_{\beta}(\boldsymbol{s}_{i},\epsilon)&=&\left(\Phi_{\beta}(\boldsymbol{s}_{i})
  +\epsilon\Phi_{\alpha}(\boldsymbol{s}_{i})\right)/N(\epsilon), \nonumber \\
N(\epsilon)&=& \sqrt{1+\epsilon^2+2\epsilon e^{-s^2_{i}/{4b^2}}},\nonumber \\
\Phi_{\alpha}(\boldsymbol{s}_{i})&=&\left(\frac{1}{\pi b^2}\right)^{\frac{3}{4}}
e^{-\frac{1}{2b^2}\left(\boldsymbol{r_\alpha}-\frac{2}{5}s_{i}\right)^2},\nonumber \\
\Phi_{\beta}(-\boldsymbol{s}_{i})&=&\left(\frac{1}{\pi b^2}\right)^{\frac{3}{4}}
e^{-\frac{1}{2b^2}\left(\boldsymbol{r_\beta}+\frac{3}{5}s_{i}\right)^2},
\end{eqnarray}
with $\boldsymbol{s}_{i}$, $i=(1,2,..., n)$, to be the generating coordinates, which are introduced to
expand the relative motion wave function~\cite{Wu:1998wu,Ping:1998si,Pang:2001xx}. The parameter $b$ indicates the size of the baryon and meson clusters, which is determined by fitting the radius of the baryon and meson by the variational method~\cite{Huang:2018rpb}. In addition, The mixing parameter $\epsilon(s_{i})$ is not an adjusted one but is determined variationally by the dynamics of the multi-quark system itself. This assumption allows the multi-quark system to choose its favourable configuration in the interacting process. It has been used to explain the crossover transition between the hadron phase\footnote{The phase shift of $NN$ interaction could be described with the formalisms with hadrons only. After including the pseudo-scalar, vector and scalar meson, especially the $\sigma$ meson, the $NN$ interaction has been well described. In Ref.~\cite{Chen:2007qn}, the authors concluded that the $\sigma$-meson exchange can be replaced by quark delocalization and color screening mechanism introduced by QDCSM by comparing the NN scattering and deuteron properties obtained by chiral quark model and QDCSM} and the quark-gluon plasma phase~\cite{Xu:2007oam, Chen:2007qn, Huang:2011kf}. Due to the effect of the mixing parameter $\epsilon(s_{i})$, there is a certain probability for the quarks between the two clusters to run, which leads to the existence of color octet states for the two clusters. Therefore, this model also includes the hidden color channel effect, which is confirmed by Refs.~\cite{Xia:2021tof, Huang:2020bmb}.

\begin{table}[htb]
\begin{center}
\renewcommand\arraystretch{1.5}
\caption{\label{channels} The relevant channels for all possible states with different $J^P$ quantum numbers}
%\resizebox{\textwidth}{!}{
\begin{tabular}{p{1.2cm}<\centering p{1.cm}<\centering p{1.cm}<\centering p{1.cm}<\centering p{1.cm}<\centering p{1.cm}<\centering p{1.cm}<\centering p{1.cm}<\centering p{1.cm}<\centering p{1.0cm}<\centering p{1.0cm}<\centering p{1.0cm}<\centering p{1.0cm}<\centering p{1.0cm}<\centering p{1.0cm}<\centering p{1.0cm}<\centering p{1.0cm}<\centering p{1.0cm}<\centering}
%\begin{tabular}{ccccc|cccc}
\toprule[1pt]
\multicolumn{7}{c}{$I=\frac{1}{2}$}\\
\midrule[1pt]
\multirow{2}{*}{$J^{P}=\frac{1}{2}^{-}$}  &$\Xi \eta_{c}$ &$\Xi J/\psi$ &$\Xi^{\ast} J/\psi$  &$\Xi_{c}^{\prime} \bar{D}_{s}$ &$\Xi_{c}\bar{D}_{s}$ &$\Xi_{c}^{\prime}\bar{D}_{s}^{\ast}$ \\
&$\Xi_{c}\bar{D}_{s}^{\ast}$ &$\Xi^{\ast}\bar{D}_{s}^{\ast}$ &$\Omega_{c}\bar{D}$ &$\Omega_{c}\bar{D}^{\ast}$ &$\Omega_{c}^{\ast}\bar{D}^{\ast}$\\
\midrule[1pt]
\multirow{2}{*}{$J^{P}=\frac{3}{2}^{-}$} &$\Xi J/\psi$ &$\Xi^{\ast} \eta_{c}$ &$\Xi^{\ast} J/\psi$ &$\Xi_{c}^{\prime} \bar{D}_{s}^{\ast}$ &$\Xi_{c} \bar{D}_{s}^{\ast}$ &$\Xi_{c}^{\ast}\bar{D}_{s}$\\
& $\Xi_{c}^{\ast}\bar{D}_{s}^{\ast}$ &$\Omega_{c}\bar{D}^{\ast}$   & $\Omega_{c}^{\ast}\bar{D}$ & $\Omega_{c}^{\ast} \bar{D}^{\ast}$\\
\midrule[1pt]
\multirow{1}{*}{$J^{P}=\frac{5}{2}^{-}$} &$\Xi^{\ast} J/\psi$ & $\Xi_{c}^{\ast}\bar{D}_{s}^{\ast}$ & $\Omega_{c}^{\ast} \bar{D}^{\ast}$\\

\bottomrule[1pt]
\end{tabular}
%}
\end{center}
\end{table}

\section{The results and discussions\label{dis}}
\begin{figure}[t]
\includegraphics[scale=0.3]{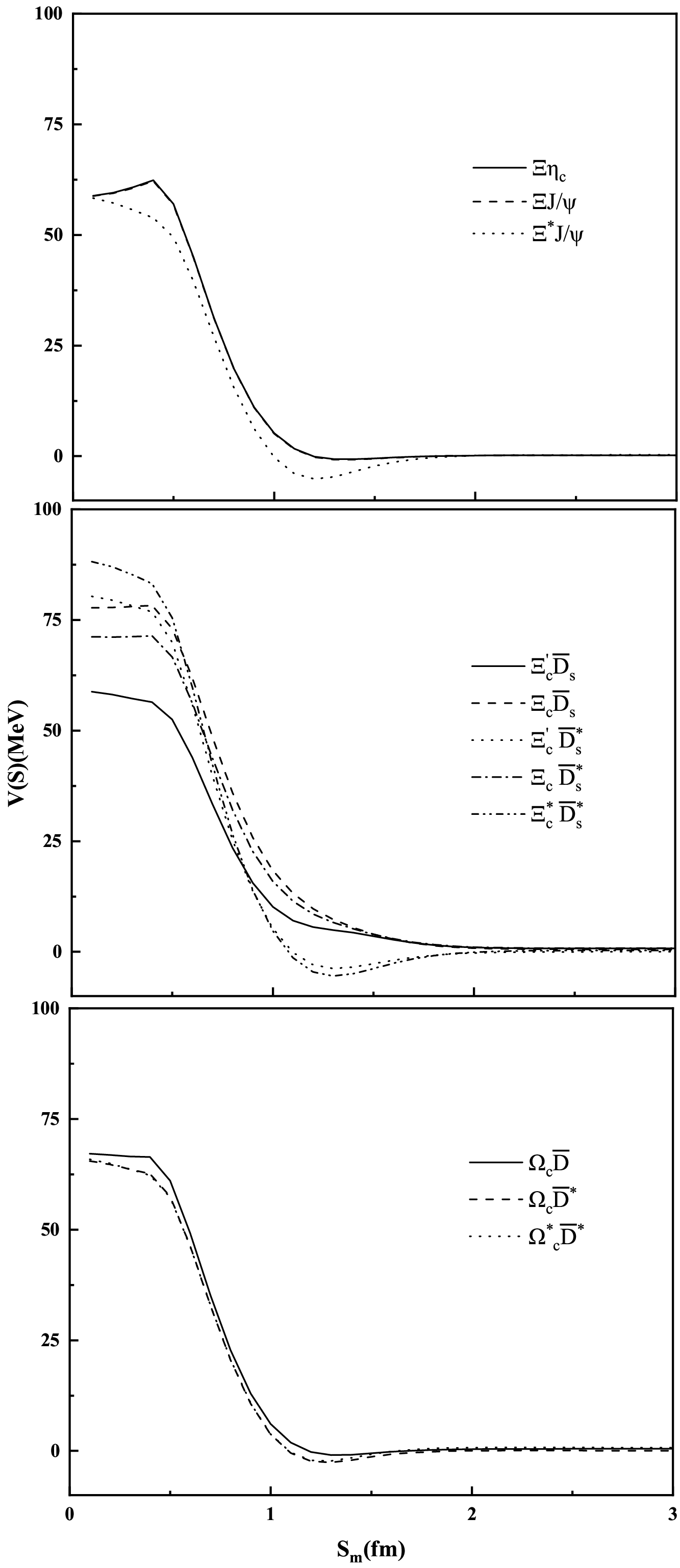}
 \caption{ The effective potentials defined in Eq.~(\ref{Eq:PotentialV}) for different channels of the hidden charm and double strangeness pentaquark systems with $J^{P}=1/2^{-}$ in QDCSM. }
\label{Veff-0.5}
\end{figure}

In this work, the low-lying pentaquark states with hidden charm and double strangeness are systematically investigated within the framework of the QDCSM. This work mainly focuses on the pentaquark states in the molecular scenario. Investigations into hidden charm and double strangeness pentaquark states are limited to the S$-$wave, restricting the quantum numbers of the system to I=1/2 and S=1/2, 3/2 and 5/2. According to the symmetry requirements of the four degrees of freedom, all possible physical channels corresponding to different quantum numbers are considered, as listed in Table~\ref{channels}. Our purpose of this work is to investigate whether possible pentaquark states with hidden charm and double strangeness exist and to see whether those pentaquark states can be explained as molecular pentaquarks. To achieve these objectives, based on the resonance group method (RGM) and generator coordinates method, the details of which can be found in Refs.~\cite{Kamimura:1977okl,wheeler1937molecular, Hill:1952jb, Griffin:1957zza, Liu:2023oyc}, the effective potentials between each physical channel are explored to predict the likelihood of bound state formation in certain channels, as depicted in Figs.~\ref{Veff-0.5}, \ref{Veff-1.5} and \ref{Veff-2.5}. Subsequently, given the potential analysis, the dynamical calculations involving single-channel and multi-channel coupling are essential to determine the existence of bound states in the $nssc\bar{c}$ system. The estimated results can be found in Tables.~\ref{bound-0.5}, \ref{bound-1.5} and \ref{bound-2.5}, respectively. Finally, to confirm whether the predicted quasi-bound states can form resonance states after coupling with certain open channels, scattering processes of these channels are examined, which can be seen in Figs.~\ref{0.5-two-coupling}, \ref{0.5-three-coupling}, \ref{0.5-1.5-two-coupling-6-9}, \ref{0.5-1.5-two-coupling-7-8}, \ref{0.5-1.5-two-coupling-10}, \ref{0.5-1.5-six-coupling} and \ref{0.5-2.5-two-coupling}.

%$CC$ stands for the effective potentials after considering the all channel coupling.

\subsection{the $J^{P}=\frac{1}{2}^{-}$}

To search the possible bound states and resonance states composed of the hadron pair listed in Table~\ref{channels}, we first estimate the effective potentials between these hadron pairs. The potential is defined as
\begin{eqnarray}
E(S_{m})&=&\frac{\langle\Psi_{5q}(S_m)|H|\Psi_{5q}(S_m)\rangle}{\langle\Psi_{5q}(S_m)|\Psi_{5q}(S_m)\rangle},\label{Eq:PotentialE}
\end{eqnarray}
where $S_m$ denotes the distance between two clusters and $\Psi_{5q}(S_m)$ represents the wave function of a certain given channel. The terms  $\langle\Psi_{5q}(S_m)|H|\Psi_{5q}(S_m)\rangle$ and $\langle\Psi_{5q}(S_m)|\Psi_{5q}(S_m)\rangle$ correspond to the Hamiltonian matrix and the overlap of the states, respectively. Thus, the effective potential between two colorless clusters is defined as
\begin{eqnarray}
	V(S_m)=E(S_m)-E(\infty), \label{Eq:PotentialV}
\end{eqnarray}
where $E(\infty)$ represents the energy at a sufficiently large distance between two clusters. The estimated potentials for $J^{P}=1/2^{-}$ are presented in Fig.~\ref{Veff-0.5}.

From the Fig.~\ref{Veff-0.5}, effective potentials of the eleven physical channels, $\Xi \eta_{c}$, $\Xi J/\psi$, $\Xi^{\ast} J/\psi$, $\Xi_{c}^{\prime} \bar{D}_{s}$, $\Xi_{c}\bar{D}_{s}$, $\Xi_{c}^{\prime}\bar{D}_{s}^{\ast}$, $\Xi_{c}\bar{D}_{s}^{\ast}$, $\Xi^{\ast}\bar{D}_{s}^{\ast}$, $\Omega_{c}\bar{D}$, $\Omega_{c}\bar{D}^{\ast}$ and $\Omega_{c}^{\ast}\bar{D}^{\ast}$, are investigated. The effective potentials of five physical channels, $\Xi \eta_{c}$, $\Xi J/\psi$, $\Xi_{c}^{\prime} \bar{D}_{s}$, $\Xi_{c}\bar{D}_{s}$ and $\Xi_{c}\bar{D}_{s}^{\ast}$, are repulsive while the remaining physical channels show attractive potential. Through analyzing these channels with effective attraction, it is evident that the effective attraction between $\Xi_{c}^{\prime} \bar{D}_{s}^{\ast}$ and $\Xi^{\ast}_{c} \bar{D}_{s}^{\ast}$ surpasses that of other channels, suggesting the $\Xi_{c}^{\prime} \bar{D}_{s}^{\ast}$ and $\Xi^{\ast}_{c} \bar{D}_{s}^{\ast}$ are highly probable to form bound states. Nonetheless, Fig.~\ref{Veff-0.5} demonstrates that the effective attraction of $\Xi^{\ast} J/\psi$, $\Omega_{c}\bar{D}$, $\Omega_{c}\bar{D}^{\ast}$ and $\Omega_{c}^{\ast}\bar{D}^{\ast}$  is relatively weak, implying that these four channels are less likely to form a bound state.

Based on the aforementioned analysis of the effective potential, it is essential to perform dynamic calculations of the bound states to confirm whether these physical channels with effective attraction form bound states.  In single-channel estimation, bound states can be identified if the lowest eigenvalues obtained fall below the theoretical thresholds of the corresponding channels. For the multi-channel coupling estimation, the lowest eigenvalue that is below the theoretical threshold of the lowest channel indicates the presence of a bound state. Notably, the current estimation is conducted in a finite space, meaning the number of basis functions is limited. Therefore, even channels with repulsive properties would still produce a series of finite basis functions in a single-channel estimation. However, as repulsive channels cannot form bound states, the eigenenergies obtained in this scenario are not the eigenvalues of bound states, and all exceed the theoretical thresholds of the respective channels. The eigenenergy of the repulsive channel would asymptotically approach the threshold of the corresponding channel when the system's space is sufficiently large due to the correlation between the eigenvalue and the number of basis functions of repulsive channels. In the case of bound states, the estimated eigenvalue is below the threshold and remains constant as the space increases. Overall, an identified bound state should have the lowest eigenvalue below the threshold of the corresponding channel and remain stable under strong decays.

The estimated results are shown in Table.~\ref{bound-0.5}. $E_{sc}$ and $E_{cc}$ denote the single channel and channel coupling estimation results, respectively, whereas $E_{th}^{Model}$ and $E_{th}^{Exp}$ represent the theoretical estimations and experimental measurements of the corresponding channel thresholds. The estimated binding energy $E_{B}=E_{sc}-E_{th}^{Model}$ can be derived from single-channel estimation. It is noteworthy that the relevant parameters used in the present research are determined according to various characteristics of different hadrons, indicating that uncertainties in the model parameters would affect the accuracy of the model predictions. To reduce the dependence on model parameters to some extent, single channel estimation correction and channel coupling estimation correction are realized by using mass splitting $E_{B}$. For instance, the corrected single channel eigenvalue $E_{sc}^{\prime}$ is calculated using $E_{B}+E_{th}^{Exp}$, while the corrected channel coupling estimation $E_{cc}^{\prime}$ is derived with reference to the lowest threshold of all channels.

Table.~\ref{bound-0.5} shows that bound states $\Xi^{\prime}_{c}\bar{D}_{s}^{\ast}$ and $\Xi_{c}^{\ast}\bar{D}_{s}^{\ast}$ with binding energies of -8 MeV and -13 MeV are obtained in the single-channel estimation. In conjunction with the analysis of the effective potential in Fig.~\ref{Veff-0.5}, the obtained bound states $\Xi^{\prime}_{c}\bar{D}_{s}^{\ast}$ and $\Xi_{c}^{\ast}\bar{D}_{s}^{\ast}$, due to strong attractive characteristics, yield lowest eigenvalues lower than the theoretical thresholds of their respective channels. However, for $\Xi^{\ast}J/\psi$, $\Omega_{c}\bar{D}$, $\Omega_{c}\bar{D}^{\ast}$ and $\Omega_{c}^{\ast} \bar{D}^{\ast}$, their weak attraction prevents the formation of bound states, with the single-channel estimated eigenvalues exceeding the theoretical thresholds of the corresponding channels, indicating that these channels transition into scattering states. For the remaining repulsive physical channels, the lowest obtained eigenvalues are higher than the corresponding physical thresholds, and thus these channels also transition into scattering states. In the multi-channel coupling estimation, Table.~\ref{bound-0.5} shows that the lowest obtained eigenvalue is about 2 MeV above the theoretical threshold of the lowest channel $\Xi \eta_{c}$, suggesting that the channel coupling effect is insufficient to produce a bound state. 

%%%%%%%%%%%%%%%%%%%%%%%%%%%%%%%%%%%%%%%%%%%%%%%%%%%%%%%%%%%%%%%%%%%%%%%%%%%%%%%%%%%%%%
Given the aforementioned effective potential and bound state estimation results analysis, although two bound states, $\Xi_{c}^{\prime} \bar{D}^{\ast}$ and $\Xi_{c}^{\ast} \bar{D}^{\ast}$, are identified through single-channel estimation, the results from Table~\ref{bound-0.5} show that the obtain lowest eigenvalues of these two bound states exceed the threshold of the lowest channel $\Xi \eta_{c}$. This implies that they may couple with specific open channels and decay through those channels. Taking the channel-coupling effects into account, these bound states might evolve into resonance states or scattering states in particular scattering process of certain open channels. To confirm this possibility, we would proceed with relevant estimation in the scattering process of certain open channels to identify the possible resonance states. 

\begin{figure*}[t]
    \includegraphics[scale=0.35]{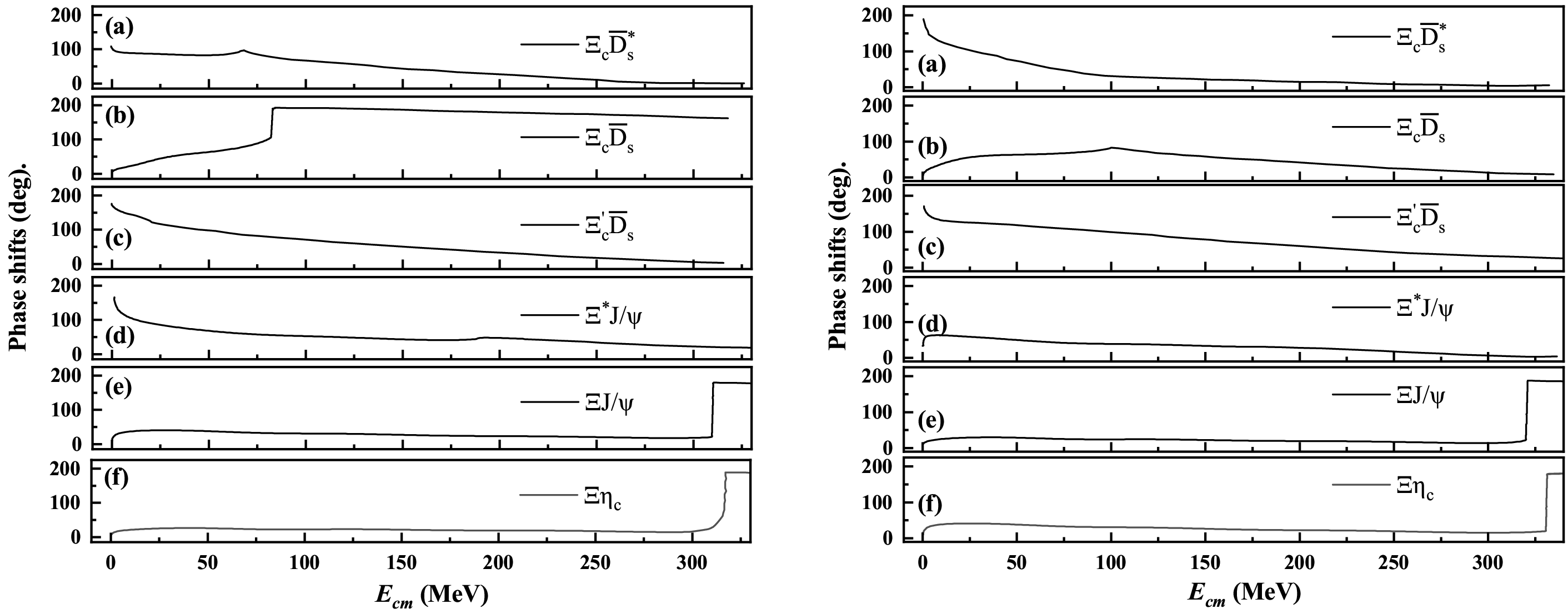}
    \caption{The phase shiftes of the open channels with two-channel coupling for $J^{P}=1/2^{-}$ in QDCSM. On the left, (a) corresponds to two-channel coupling with $\Xi_{c}^{\prime}\bar{D}_{s}^{\ast}$ and $\Xi_{c}\bar{D}_{s}^{\ast}$, (b) denotes two-channel coupling with $\Xi_{c}^{\prime}\bar{D}_{s}^{\ast}$ and $\Xi_{c}\bar{D}_{s}$, (c) shows two-channel coupling with $\Xi_{c}^{\prime}\bar{D}_{s}^{\ast}$ and $\Xi_{c}^{\prime}\bar{D}_{s}$, (d) stands for two-channel coupling with $\Xi_{c}^{\prime}\bar{D}_{s}^{\ast}$ and $\Xi^{\ast} J/\psi$, (e) represents for two-channel coupling with $\Xi_{c}^{\prime}\bar{D}_{s}^{\ast}$ and $\Xi J/\psi$, (f) represents for two-channel coupling with $\Xi_{c}^{\prime}\bar{D}_{s}^{\ast}$ and $\Xi \eta_{c}$. On the right, (a) corresponds to two-channel coupling with $\Xi_{c}^{\ast}\bar{D}_{s}^{\ast}$ and $\Xi_{c}\bar{D}_{s}^{\ast}$, (b) denotes two-channel coupling with $\Xi_{c}^{\ast}\bar{D}_{s}^{\ast}$ and $\Xi_{c}\bar{D}_{s}$, (c) shows two-channel coupling with $\Xi_{c}^{\ast}\bar{D}_{s}^{\ast}$ and $\Xi_{c}^{\prime}\bar{D}_{s}$, (d) stands for two-channel coupling with $\Xi_{c}^{\ast}\bar{D}_{s}^{\ast}$ and $\Xi^{\ast} J/\psi$, (e) represents for two-channel coupling with $\Xi_{c}^{\ast}\bar{D}_{s}^{\ast}$ and $\Xi J/\psi$, (f) represents for two-channel coupling with $\Xi_{c}^{\ast}\bar{D}_{s}^{\ast}$ and $\Xi \eta_{c}$.}
    \label{0.5-two-coupling}
\end{figure*}

Here, two types of channel coupling are utilized to assess the influence of various channel coupling effects on the formation of resonance states. The first method examines the two-channel coupling of one bound state with the corresponding open channel, while the second approach couples all bound states with a specific open channel. The two-channel coupling scenario involving one bound state and its corresponding open channel is first analyzed. As illustrated in Fig.~\ref{0.5-two-coupling}, the behavior of bound states $\Xi_{c}^{\prime} \bar{D}^{\ast}$ and $\Xi_{c}^{\ast} \bar{D}^{\ast}$ in different open channel scattering processes is displayed. If a sudden 180-degree phase shift surge is observed when the incident energy increases, indicating that certain bound states have transitioned into resonance states; otherwise, they would convert into scattering states. From Fig.~\ref{0.5-two-coupling}, it is visually apparent the bound states $\Xi_{c}^{\prime} \bar{D}^{\ast}$ and $\Xi_{c}^{\ast} \bar{D}^{\ast}$ can decay into open channels  $\Xi \eta_{c}$, $\Xi J/\psi$, $\Xi^{\ast} J/\psi$, $\Xi^{\prime}_{c} \bar{D}_{s}$, $\Xi_{c} \bar{D}_{s}$, and $\Xi_{c} \bar{D}_{s}^{\ast}$, respectively. For the bound state $\Xi_{c}^{\prime} \bar{D}^{\ast}$, the resonance state $\Xi_{c}^{\prime} \bar{D}^{\ast}$  is detected in the scattering phase shifts of open channels $\Xi \eta_{c}$, $\Xi J/\psi$ and $\Xi_{c}\bar{D}_{s}$, while no similar signal is observed in the other open channels. In the case of the bound state $\Xi_{c}^{\ast} \bar{D}_{s}^{\ast}$, a 180-degree phase shifts surge is observed in the scattering processes of open channels $\Xi \eta_{c}$ and $\Xi J/\psi$, confirming the existence of resonance state $\Xi_{c}^{\ast} \bar{D}_{s}^{\ast}$; however, no similar behaviors are found in scattering processes of open channels $\Xi^{\ast} J/\psi$, $\Xi^{\prime}_{c} \bar{D}_{s}$, $\Xi_{c} \bar{D}_{s}$, and $\Xi \bar{D}_{s}^{\ast}$. 

\begin{figure*}[t]
    \includegraphics[scale=0.35]{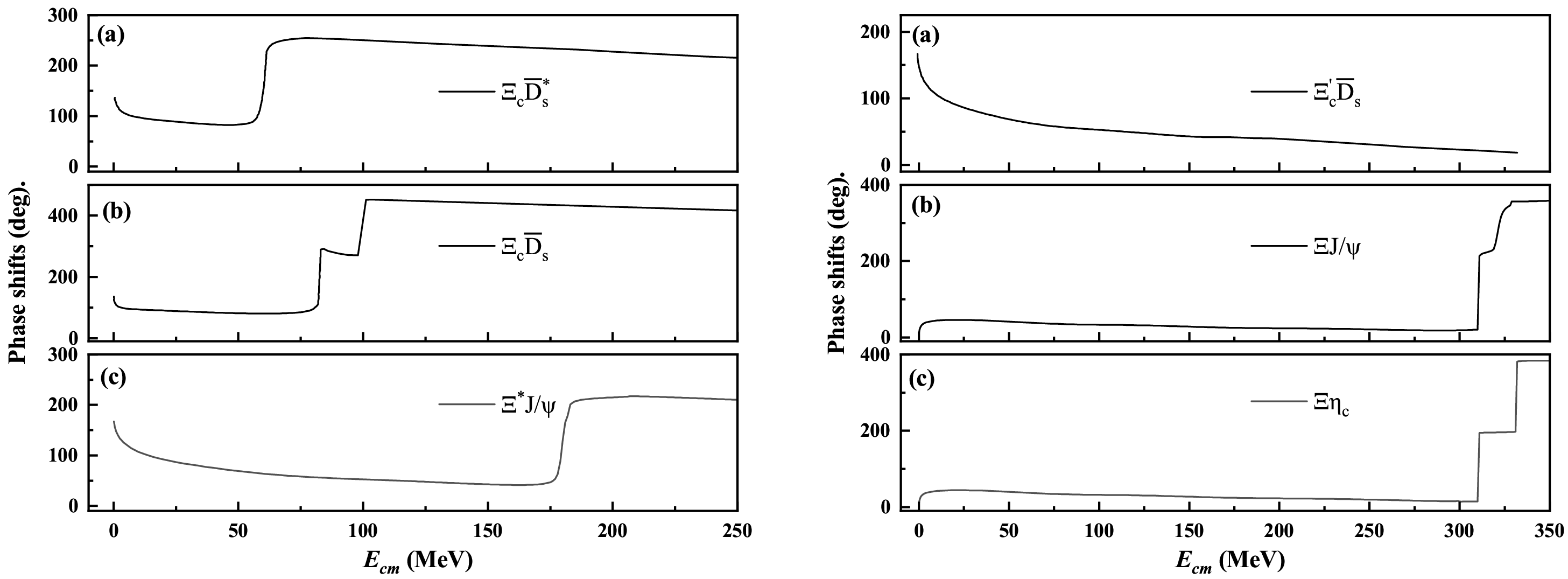}
    \caption{The phase shifts of the open channels with three-channel coupling with two closed channels ($\Xi_{c}^{\ast}\bar{D}_{s}^{\ast}$ and $\Xi_{c}^{\prime}\bar{D}_{s}^{\ast}$) and one open channel for  $J^{P}=1/2^{-}$ in QDCSM. On the left, (a) stands for three-channel coupling with $\Xi_{c}^{\ast}\bar{D}_{s}^{\ast}$, $\Xi_{c}^{\prime}\bar{D}_{s}^{\ast}$ and $\Xi_{c} \bar{D}_{s}^{\ast}$, (b) represents three-channel coupling with $\Xi_{c}^{\ast}\bar{D}_{s}^{\ast}$, $\Xi_{c}^{\prime}\bar{D}_{s}^{\ast}$ and $\Xi_{c} \bar{D}_{s}$, (c) denotes three-channel coupling with $\Xi_{c}^{\ast}\bar{D}_{s}^{\ast}$, $\Xi_{c}^{\prime}\bar{D}_{s}^{\ast}$ and $\Xi^{\ast} J/\psi$. On the right, (a) shows three-channel coupling with $\Xi_{c}^{\ast}\bar{D}_{s}^{\ast}$, $\Xi_{c}^{\prime}\bar{D}_{s}^{\ast}$ and $\Xi_{c}^{\prime} \bar{D}_{s}$, (b) represents three-channel coupling with $\Xi_{c}^{\ast}\bar{D}_{s}^{\ast}$, $\Xi_{c}^{\prime}\bar{D}_{s}^{\ast}$ and $\Xi J/\psi$, (c) denotes three-channel coupling with $\Xi_{c}^{\ast}\bar{D}_{s}^{\ast}$, $\Xi_{c}^{\prime}\bar{D}_{s}^{\ast}$ and $\Xi \eta_{c}$.}
    \label{0.5-three-coupling}
\end{figure*}

Subsequently, the three-channel coupling case of two bound states and one open channel is examined. As shown in Fig.~\ref{0.5-three-coupling}, two resonance states, $\Xi_{c}^{\prime} \bar{D}^{\ast}$ and $\Xi_{c}^{\ast} \bar{D}^{\ast}$, are simultaneously observed in the scattering processes of open channels $\Xi \eta_{c}$, $\Xi J/\psi$, and $\Xi_{c} \bar{D}_{s}$. However, only one resonance state $\Xi_{c}^{\prime} \bar{D}^{\ast}$ is derived in the scattering processes of open channels  $\Xi^{\ast} J/\psi$  and $\Xi_{c} \bar{D}_{s}^{\ast}$, while resonance state $\Xi_{c}^{\ast} \bar{D}^{\ast}$ disappears in the corresponding scattering processes after the three-channel coupling estimation. This occurs because the channel-coupling estimation raises the energy of the higher bound states $\Xi_{c}^{\ast} \bar{D}^{\ast}$ beyond its theoretical threshold, causing bound state $\Xi_{c}^{\ast} \bar{D}^{\ast}$  to transform into a scattering state in the scattering processes of open channels $\Xi^{\ast} J/\psi$  and $\Xi_{c} \bar{D}_{s}^{\ast}$. A similar phenomenon can be detected in the scattering process of open channel $\Xi^{\prime}_{c} \bar{D}_{s}$, with the only difference being that both bound states $\Xi_{c}^{\prime} \bar{D}^{\ast}$ and $\Xi_{c}^{\ast} \bar{D}^{\ast}$ transform into scattering states. 

Therefore, Table.~\ref{phaseshifts-0.5} summarizes the masses and widths of the resonance states obtained from Fig.~\ref{0.5-two-coupling} and~\ref{0.5-three-coupling}. According to Table.~\ref{phaseshifts-0.5}, the resonance mass $M_{R}^{th}$ is derived by summing the threshold of the corresponding open channel with the incident energy at a scattering phase shift of $\pi/2$, while $M_{R}$ indicates the corrected resonance mass, calculated via $M_{R}=M_{R}^{th}-\sum_{i}p_{i}(E_{th}^{Model}(i)-E_{th}^{Exp}(i))$, where $p_{i}$ denotes the contribution of the $i$th channel in the channel coupling estimation. The decay width of resonance states observed in different open channel scattering processes can be determined by computing the difference in incident energy corresponding to phase shifts between $\pi/4$ and $3\pi/4$. Table.~\ref{phaseshifts-0.5} shows that the corrected mass range of resonance $\Xi_{c}^{\prime} \bar{D}^{\ast}$ is approximately $4682$ MeV to $4688$ MeV, with a decay width range of about $6.4$ MeV to $24.2$ MeV. The other resonance state $\Xi_{c}^{\ast} \bar{D}^{\ast}$ has a corrected mass of $4751$ MeV$-$4756 MeV and a decay width of 3.2 MeV$-$18.7 MeV.

%%%%%%%%%%%%%%%%%%%%%%%%%%%%%%%%%%%%%%%%%%%%%%%%%%%%%%%%%%%%%%%%%%%%%%%%%%%%%%%%%%%%%%%%%%%%%%%%%%%%%%%%%%%%%%%%%%%%%%%%%%%%%%%%%%%%%%%%%%%

\begin{table}[htb]
    \begin{center}
        \caption{\label{bound-0.5}The binding energies and the masses of every single channel and those of channel coupling for the hidden charm pentaquarks with double strangeness with $J^{P}=1/2^{-}$. The values are provided in units of MeV. }
        \renewcommand\arraystretch{1.5}
        \resizebox{0.48\textwidth}{!} {
        \begin{tabular}{p{1.cm}<\centering p{1.cm}<\centering p{1.cm}<\centering p{1.cm}<\centering p{1.cm}<\centering p{1.cm}<\centering p{1.cm}<\centering p{1.cm}<\centering p{1.0cm}}
            %\begin{tabular}{ccccc|cccc}
            \toprule[1pt]
            
            Channel    & $E_{sc}$  &$E_{th}^{Model}$   &$E_{B}$  &$E_{th}^{Exp}$  &$E_{sc}^{\prime}$  &$E_{cc}/E_{B}$  & $E_{cc}^{\prime}$ \\
            \midrule[1pt]
            $\Xi \eta_{c}$                            &4362  &4360 &2  &4298  &4300  &\multirow{11}{*}{4362/2}   &\multirow{11}{*}{4300}  \\
            $\Xi J/\psi$                              &4365  &4363 &2  &4414  &4416  \\
            $\Xi^{\ast} J/\psi$                       &4486  &4485 &1  &4629  &4630 \\                 
            $\Xi_{c}^{\prime} \bar{D}_{s}$            &4640  &4638 &2  &4545  &4547\\
            $\Xi_{c} \bar{D}_{s}$                     &4571  &4569 &2  &4435  &4437\\
            $\Xi_{c}^{\prime} \bar{D}_{s}^{\ast}$     &4644  &4652 &-8 &4689  &4681 \\
            $\Xi_{c} \bar{D}_{s}^{\ast}$              &4585  &4583 &2  &4579  &4581\\
            $\Xi_{c}^{\ast} \bar{D}_{s}^{\ast}$       &4658  &4671 &-13&4757  &4744 \\
            $\Omega_{c} \bar{D}$                      &4672  &4671 &1  &4564  &4565\\
            $\Omega_{c} \bar{D}^{\ast}$               &4706  &4705 &1  &4702  &4703\\
            $\Omega_{c}^{\ast} \bar{D}^{\ast}$        &4717  &4716 &1  &4777  &4778\\                          
            
            \bottomrule[1pt]
        \end{tabular}
        }
    \end{center}
\end{table}

\begin{table*}[htb]
    \begin{center}
        \caption{\label{phaseshifts-0.5} The masses and decay widths (in the unit of MeV) of resonance states with the difference scattering process with $J^{P}=1/2$.  }
        \renewcommand\arraystretch{1.5}
        \begin{tabular}{p{2.3cm}<\centering p{1.cm}<\centering p{1.cm}<\centering p{0.5cm}<\centering p{0.01cm}<\centering p{1.cm}<\centering p{1.cm}<\centering p{0.5cm}<\centering p{0.01cm}<\centering p{1.cm}<\centering p{1.cm}<\centering p{0.5cm}<\centering p{0.01cm}<\centering p{1.cm}<\centering p{1.cm}<\centering p{0.5cm}<\centering p{0.2cm}<\centering p{1.cm}<\centering p{1.cm}<\centering p{2.cm}<\centering p{1.cm}<\centering p{1.cm}<\centering p{1.cm}<\centering   p{2.0cm}<\centering  }
           \toprule[1pt]
           \multirow{4}{*}{} & \multicolumn{6}{c}{Two-channel coupling} &  & \multicolumn{6}{c}{Three-channel coupling} \\
           \cline{2-8} \cline{10-16}

           &\multicolumn{3}{c}{$\Xi_{c}^{\prime} \bar{D}_{s}^{\ast}$} &  &\multicolumn{3}{c}{$\Xi_{c}^{\ast} \bar{D}_{s}^{\ast}$} &  &\multicolumn{3}{c}{$\Xi_{c}^{\prime} \bar{D}_{s}^{\ast}$} &  &\multicolumn{3}{c}{$\Xi_{c}^{\ast} \bar{D}_{s}^{\ast}$}\\
           \cline{2-4}\cline{6-8}\cline{10-12}\cline{14-16}
           Open channels & $M_{R}^{th}$ & $M_{R}$ & $\Gamma_{i}$  &     &$M_{R}^{th}$ & $M_{R}$ &$\Gamma_{i}$      & & $M_{R}^{th}$ & $M_{R}$ & $\Gamma_{i}$  &  & $M_{R}^{th}$ & $M_{R}$ & $\Gamma_{i}$  \\
           \midrule[1pt]
           $\Xi \eta_{c}$      &4650  &4687     &2.5  &   &4668   &4754  &2.1 &  &4649 &4686   &1.4  & &4669      &4755   &1.9  & \\
           $\Xi J/\psi$        &4651  &4688     &2.2  &   &4667   &4753  &1.1 &  &4650 &4687   &1.3  & &4665      &4751   &1.0  & \\
           $\Xi^{\ast} J/\psi$ &$\ldots$ &$\ldots$ &$\ldots$ & &$\ldots$ &$\ldots$ &$\ldots$  & &4645 & 4682 &12.5& &$\ldots$ &$\ldots$ &$\ldots$ &\\
           $\Xi^{\prime}_{c} \bar{D}_{s}$&$\ldots$ &$\ldots$ &$\ldots$ & &$\ldots$ &$\ldots$ &$\ldots$ & &$\ldots$ &$\ldots$ &$\ldots$ & &$\ldots$ &$\ldots$ &$\ldots$ & \\
           $\Xi_{c} \bar{D}_{s}$ &4651  &4688 &1.7 & &$\ldots$ &$\ldots$ &$\ldots$ & &4651 &4688 &1.5 & &4670 &4756 &15.8 & \\
           $\Xi_{c} \bar{D}_{s}^{\ast}$ &$\ldots$ &$\ldots$ &$\ldots$ & &$\ldots$ &$\ldots$ &$\ldots$ & &4651 &4688 &7.5 & &$\ldots$ &$\ldots$ &$\ldots$\\
           $\Gamma_{Total}$   & & &6.4 & & & &3.2 &&&&24.2 &&&&18.7 \\  
          \bottomrule[1pt]
        \end{tabular}
    \end{center}
\end{table*}

\subsection{the $J^{P}=\frac{3}{2}^{-}$}
\begin{figure}[t]
     \includegraphics[scale=0.3]{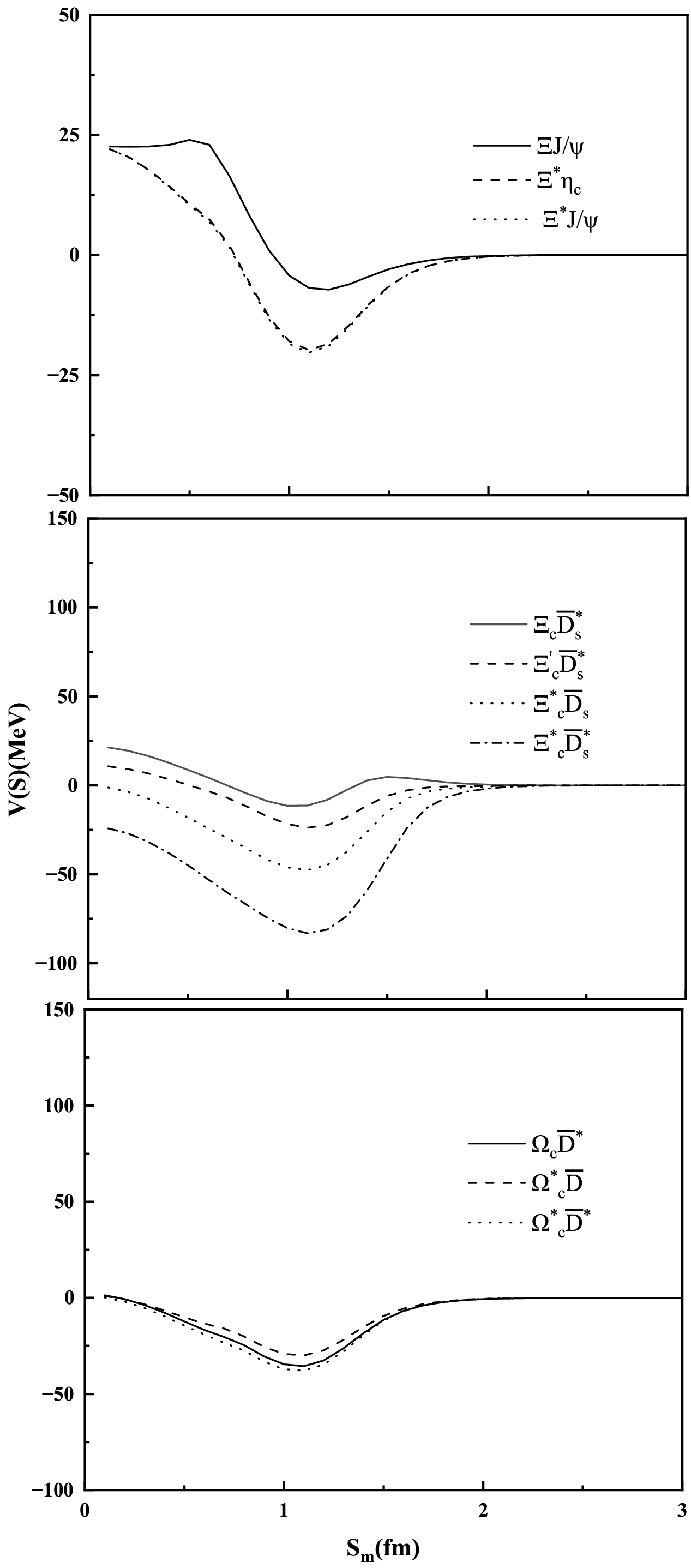}
    \caption{ The effective potentials defined in Eq.~(\ref{Eq:PotentialV}) for different channels of the hidden charm and double strangeness pentaquark systems with $J^{P}=3/2^{-}$ in QDCSM. }
    \label{Veff-1.5}
\end{figure}

%%%%%%%%%%%%%%%%%%%%%%%%%%%%%%%%%%%%%%%%%%%%%%%%%%%%%%%%%%%%%%%%%%%%%%%%%%%%%%%%%%%%%%%%%%%%%%%%%%%%%%%%%%%%%%%%%%%%%%
The behavior of $J^{P}=3/2^{-}$ system is similar to that of $J^{P}=1/2^{-}$ system, with this system encompassing ten physical channels, labeled $\Xi J/\psi$, $\Xi^{\ast} \eta_{c}$, $\Xi^{\ast} J/\psi$, $\Xi_{c}^{\prime} \bar{D}_{s}^{\ast}$,  $\Xi_{c} \bar{D}_{s}^{\ast}$,  $\Xi_{c}^{\ast} \bar{D}_{s}$, $\Xi_{c}^{\ast} \bar{D}_{s}^{\ast}$, $\Omega_{c} \bar{D}^{\ast}$, $\Omega_{c}^{\ast} \bar{D}$, and $\Omega_{c}^{\ast} \bar{D}^{\ast}$. Fig.~\ref{Veff-1.5} illustrates the trend of the effective potentials for the ten physical channels in the $J^{P}=3/2^{-}$ system, revealing that all channels exhibit attractive potentials, indicating a likelihood of forming bound states in dynamic bound-state estimation. A comparison of the channels' attractive potentials reveals that the attractive potential of $\Xi_{c}^{\ast} \bar{D}_{s}^{\ast}$  channel is significantly stronger than that of the others. Meanwhile, the attractive potentials of $\Xi_{c}^{\ast} \bar{D}_{s}$, $\Omega_{c} \bar{D}^{\ast}$, $\Omega_{c}^{\ast} \bar{D}$, and $\Omega_{c}^{\ast} \bar{D}^{\ast}$ channels show minimal variation and are consistently stronger than those of $\Xi J/\psi$, $\Xi^{\ast} \eta_{c}$, $\Xi^{\ast} J/\psi$, $\Xi_{c}^{\prime} \bar{D}_{s}^{\ast}$ and $\Xi_{c} \bar{D}_{s}^{\ast}$. This suggests that  $\Xi_{c}^{\ast} \bar{D}_{s}$, $\Omega_{c} \bar{D}^{\ast}$, $\Omega_{c}^{\ast} \bar{D}$, and $\Omega_{c}^{\ast} \bar{D}^{\ast}$ channels are more likely to form bound states in dynamic bound-state estimation.   

Table~\ref{bound-0.5} lists that lowest eigenvalues estimated for single-channel and all-channel coupling in the sector with a quantum number of $3/2^{-}$, As shown in Table~\ref{bound-0.5}, it can be observed that the single-channel estimations identifies five bound states, labeled $\Xi_{c}^{\ast} \bar{D}_{s}$, $\Xi_{c}^{\ast} \bar{D}_{s}^{\ast}$, $\Omega_{c} \bar{D}^{\ast}$, $\Omega_{c}^{\ast} \bar{D}$ and 
$\Omega_{c}^{\ast} \bar{D}^{\ast}$, respectively. Compared to the other bound states, bound state $\Xi_{c}^{\ast} \bar{D}_{s}^{\ast}$ has the highest binding energy, approximately -34 MeV, which aligns with the behavior of its effective potential. Compared with the results of Ref.~\cite{Ortega:2022uyu}, which similarly identified bound states $\Xi_{c}^{\ast} \bar{D}_{s}^{\ast} $ and $\Xi_{c}^{\ast} \bar{D}_{s}$ through single-channel estimations, this finding aligns well with the results of the present study. In Refs.~\cite{Marse-Valera:2022khy, Marse-Valera:2024apc,Magas:2024biu}, two poles were spotted, one of which exhibits the characteristics of spin-parity $1/2^{-}$ and $3/2^{-}$ and is considered to correspond to a molecular state $\Omega_{c}\bar{D}^{\ast}$. This result is consistent with the estimation of the bound state $\Omega_{c}\bar{D}^{\ast}$ obtained in the present study. Similar results can be also found in Ref.~\cite{Roca:2024nsi}. Furthermore, the estimations for all-channel coupling are derived from calculations of dynamically bound states. After accounting for the effects of full-channel coupling, the corrected lowest eigenvalue is 4415 MeV, which approximates the threshold of the lowest physical channel $\Xi J/\psi$As a result, no bound state is captured in the sector with a quantum number of $3/2^{-}$.  As a result, no bound state is captured in the sector with a quantum number of $3/2^{-}$. 

\begin{figure*}[t]
    \includegraphics[scale=0.35]{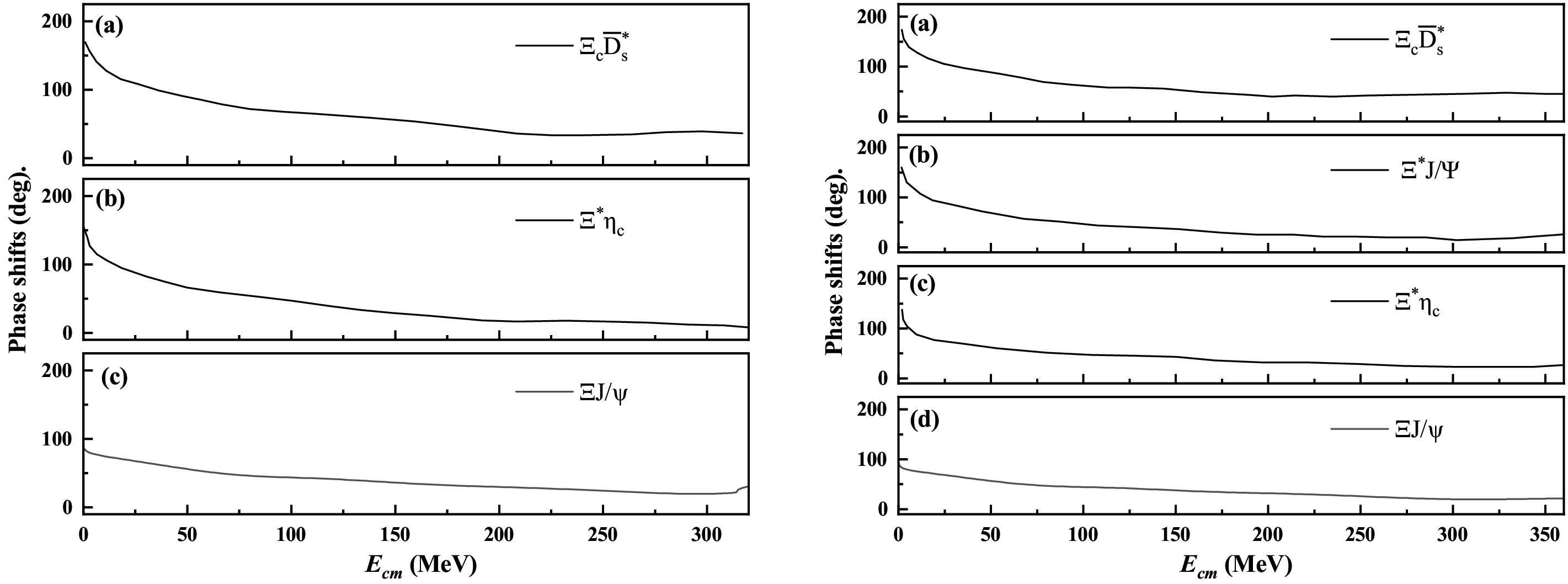}
    \caption{The phase shifts of the open channels with two-channel coupling for $J^{P}=3/2^{-}$ in QDCSM. On the left, (a) corresponds to two-channel coupling with $\Xi_{c}^{\ast}\bar{D}_{s}$ and $\Xi_{c}\bar{D}_{s}^{\ast}$, (b) denotes two-channel coupling with $\Xi_{c}^{\ast}\bar{D}_{s}$ and $\Xi^{\ast}\eta_{c}$, (c) shows two-channel coupling with $\Xi_{c}^{\ast}\bar{D}_{s}$ and $\Xi J/\psi$. On the right, (a) corresponds to two-channel coupling with $\Omega_{c}^{\ast}\bar{D}$ and $\Xi_{c}\bar{D}_{s}^{\ast}$, (b) denotes two-channel coupling with $\Omega_{c}^{\ast}\bar{D}$ and $\Xi^{\ast} J/\psi$, (c) shows two-channel coupling with $\Omega_{c}^{\ast}\bar{D}$ and $\Xi^{\ast}\eta_{c}$, (d) stands for two-channel coupling with $\Omega_{c}^{\ast}\bar{D}$ and $\Xi J/\psi$.}
    \label{0.5-1.5-two-coupling-6-9}
\end{figure*}

\begin{figure*}[t]
    \includegraphics[scale=0.35]{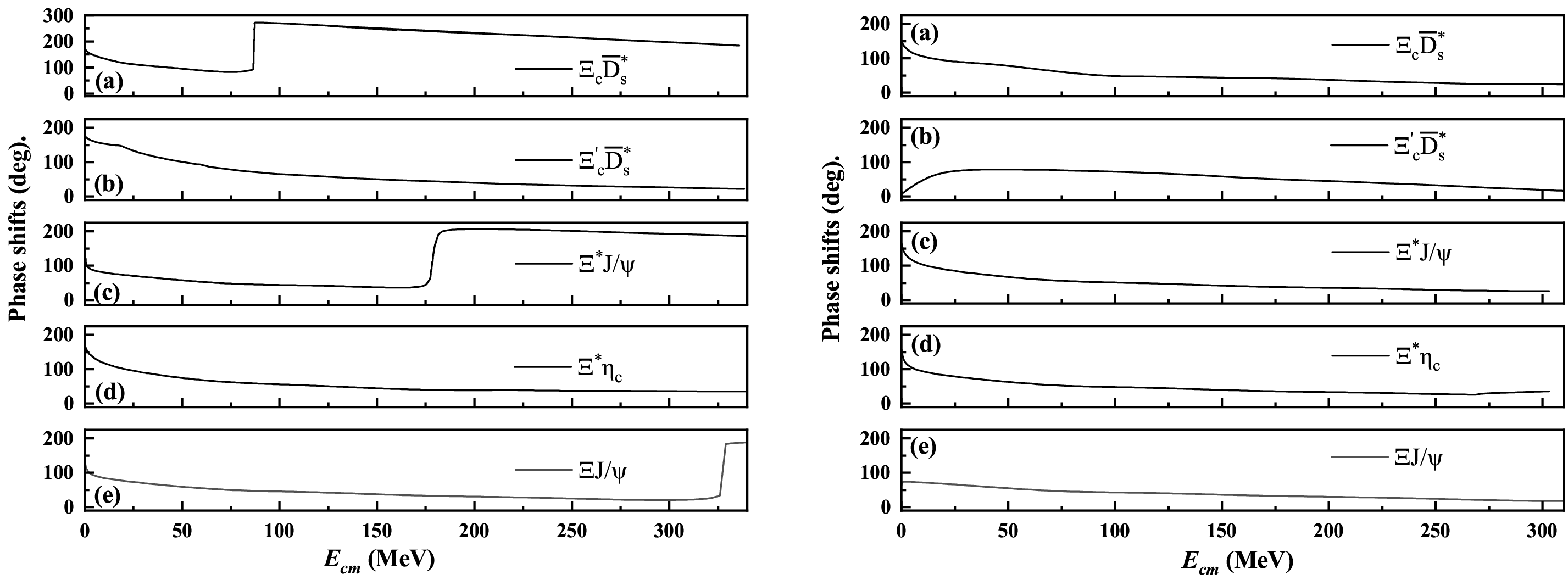}
    \caption{The phase shifts of the open channels with two-channel coupling for $J^{P}=3/2^{-}$ in QDCSM. On the left, (a) corresponds to two-channel coupling with $\Xi_{c}^{\ast}\bar{D}_{s}^{\ast}$ and $\Xi_{c}\bar{D}_{s}^{\ast}$, (b) denotes two-channel coupling with $\Xi_{c}^{\ast}\bar{D}_{s}^{\ast}$ and $\Xi_{c}^{\prime}\bar{D}_{s}^{\ast}$, (c) shows two-channel coupling with $\Xi_{c}^{\ast}\bar{D}_{s}^{\ast}$ and $\Xi^{\ast} J/\psi$, (d) stands for two-channel coupling with $\Xi_{c}^{\ast}\bar{D}_{s}^{\ast}$ and $\Xi^{\ast} \eta_{c}$, (e) represents for two-channel coupling with $\Xi_{c}^{\ast}\bar{D}_{s}^{\ast}$ and $\Xi J/\psi$. On the right, (a) corresponds to two-channel coupling with $\Omega_{c}\bar{D}^{\ast}$ and $\Xi_{c}\bar{D}_{s}^{\ast}$, (b) denotes two-channel coupling with $\Omega_{c}\bar{D}^{\ast}$ and $\Xi_{c}^{\prime}\bar{D}_{s}^{\ast}$, (c) shows two-channel coupling with $\Omega_{c}\bar{D}^{\ast}$ and $\Xi^{\ast} J/\psi$, (d) stands for two-channel coupling with $\Omega_{c}\bar{D}^{\ast}$ and $\Xi^{\ast} \eta_{c}$, (e) represents for two-channel coupling with $\Omega_{c}\bar{D}^{\ast}$ and $\Xi J/\psi$.}
    \label{0.5-1.5-two-coupling-7-8}
\end{figure*}

\begin{figure}[htp]
    \includegraphics[scale=0.35]{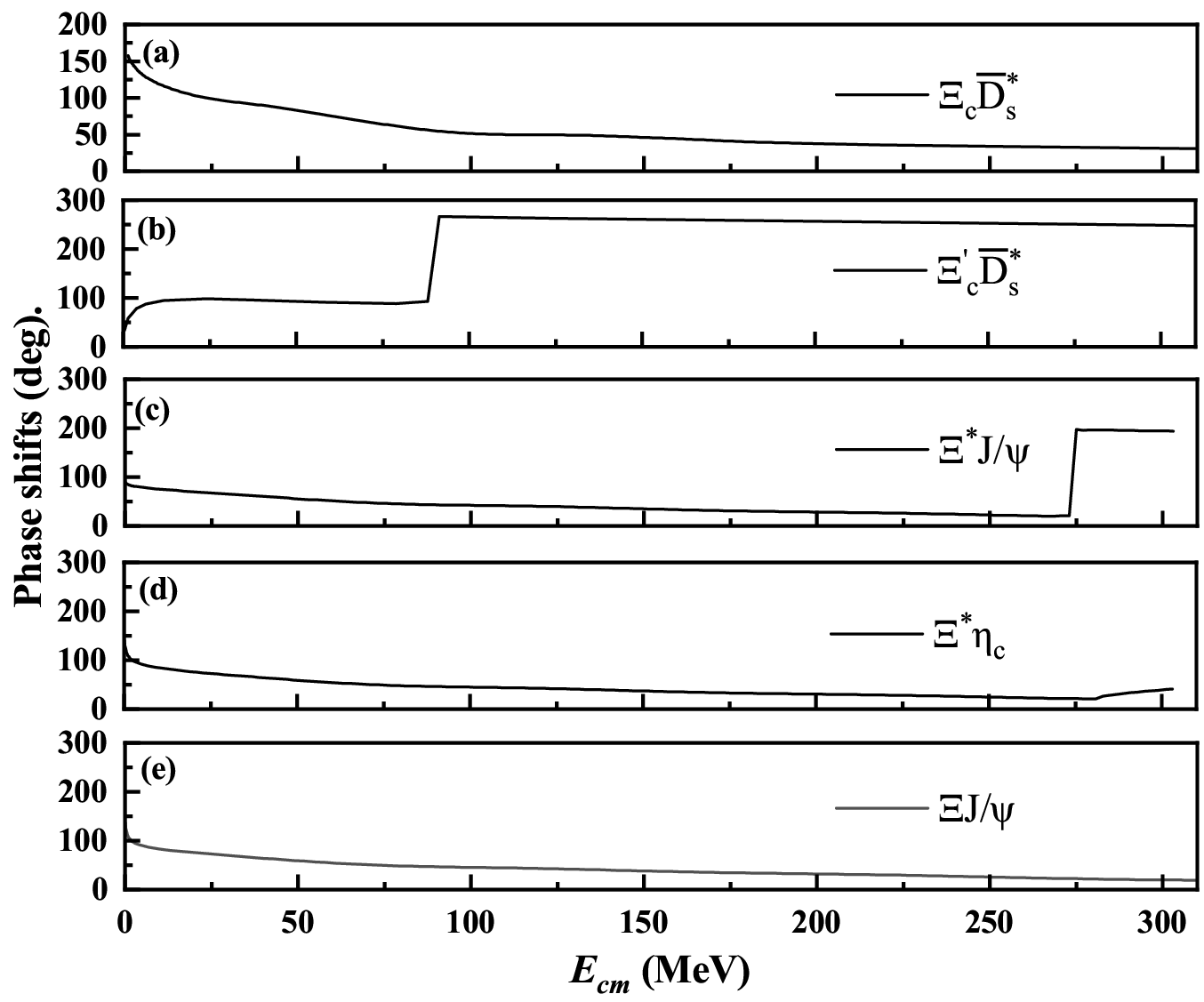}
    \caption{The phase shifts of the open channels with two-channel coupling for $J^{P}=3/2^{-}$ in QDCSM; (a) corresponds to two-channel coupling with $\Omega_{c}^{\ast}\bar{D}^{\ast}$ and $\Xi_{c}\bar{D}_{s}^{\ast}$, (b) denotes two-channel coupling with $\Omega_{c}^{\ast}\bar{D}^{\ast}$ and $\Xi_{c}^{\prime}\bar{D}_{s}^{\ast}$, (c) shows two-channel coupling with $\Omega_{c}^{\ast}\bar{D}^{\ast}$ and $\Xi^{\ast} J/\psi$, (d) stands for two-channel coupling with $\Omega_{c}^{\ast}\bar{D}^{\ast}$ and $\Xi^{\ast} \eta_{c}$, (e) represents for two-channel coupling with $\Omega_{c}^{\ast}\bar{D}^{\ast}$ and $\Xi J/\psi$.}
    \label{0.5-1.5-two-coupling-10}
\end{figure}

\begin{figure}[htp]
    \includegraphics[scale=0.35]{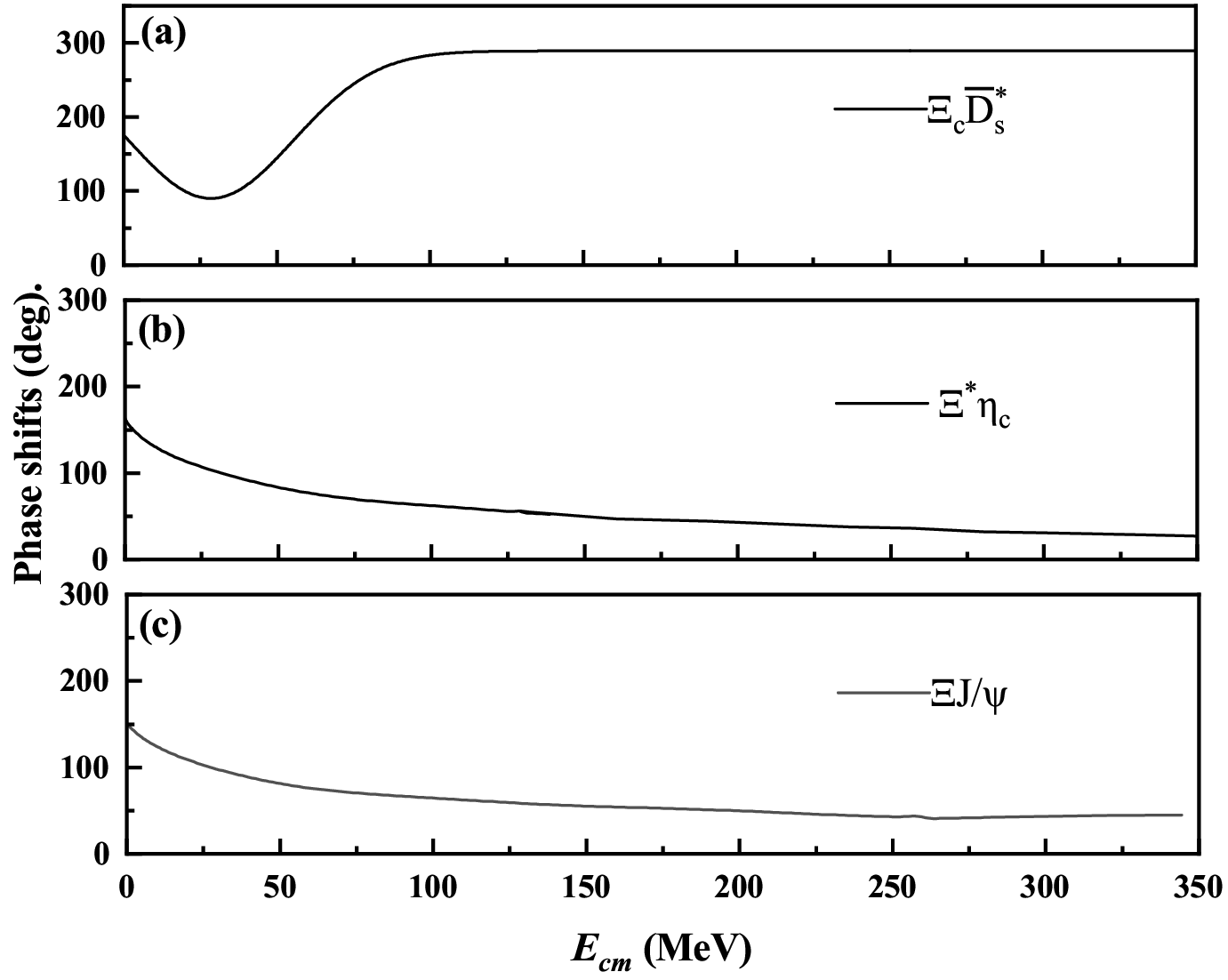}
    \caption{The phase shifts of the open channels with six-channel coupling with five closed channels ($\Xi_{c}^{\ast}\bar{D}_{s}$, $\Xi_{c}^{\ast}\bar{D}_{s}^{\ast}$, $\Omega_{c}\bar{D}^{\ast}$, $\Omega_{c}^{\ast}\bar{D}$ and $\Omega_{c}^{\ast}\bar{D}^{\ast}$) and one open channel for $J^{P}=3/2^{-}$ in QDCSM; (a) corresponds to six-channel coupling with $\Xi_{c}^{\ast}\bar{D}_{s}$, $\Xi_{c}^{\ast}\bar{D}_{s}^{\ast}$, $\Omega_{c}\bar{D}^{\ast}$, $\Omega_{c}^{\ast}\bar{D}$, $\Omega_{c}^{\ast}\bar{D}^{\ast}$, and $\Xi_{c}\bar{D}_{s}^{\ast}$,  (b) stands for six-channel coupling with $\Xi_{c}^{\ast}\bar{D}_{s}$, $\Xi_{c}^{\ast}\bar{D}_{s}^{\ast}$, $\Omega_{c}\bar{D}^{\ast}$, $\Omega_{c}^{\ast}\bar{D}$, $\Omega_{c}^{\ast}\bar{D}^{\ast}$, and $\Xi^{\ast} \eta_{c}$, (c) represents for six-channel coupling with $\Xi_{c}^{\ast}\bar{D}_{s}$, $\Xi_{c}^{\ast}\bar{D}_{s}^{\ast}$, $\Omega_{c}\bar{D}^{\ast}$, $\Omega_{c}^{\ast}\bar{D}$, $\Omega_{c}^{\ast}\bar{D}^{\ast}$, and $\Xi J/\psi$.}
    \label{0.5-1.5-six-coupling}
\end{figure}

The dynamic bound states estimation confirms the existence of five bound states. To further investigate whether these bound states can form resonance states, it is necessary to examine their scattering processes in specific open channels. The scattering behavior of these five bound states in different open channels is similar to that observed in the sector with $J^{P}=1/2^{-}$. Initially, the two-channel coupling scattering phase shifts processes involving a bound state and its corresponding open channel are studied. Fig.~\ref{0.5-1.5-two-coupling-6-9}, \ref{0.5-1.5-two-coupling-7-8} and \ref{0.5-1.5-two-coupling-10} show that bound states $\Xi_{c}^{\ast} \bar{D}_{s}^{\ast}$, $\Omega_{c} \bar{D}^{\ast}$ and $\Omega_{c}^{\ast} \bar{D}^{\ast}$ can decay into open channels $\Xi J/\psi$, $\Xi^{\ast} \eta_{c}$, $\Xi^{\ast} J/\psi$, $\Xi_{c}^{\prime} \bar{D}_{s}^{\ast}$ and $\Xi_{c} \bar{D}_{s}^{\ast}$. Resonance state $\Xi_{c}^{\ast} \bar{D}_{s}^{\ast}$ is detected in the scattering processes of open channels $\Xi J/\psi$,  $\Xi^{\ast} J/\psi$ and 
$\Xi_{c} \bar{D}_{s}^{\ast}$; resonance state $\Omega_{c}^{\ast} \bar{D}^{\ast}$ is observed in open channels  $\Xi^{\ast} J/\psi$ and $\Xi_{c}^{\prime} \bar{D}_{s}^{\ast}$, while no significant peaks for resonance states  $\Xi_{c}^{\ast} \bar{D}_{s}^{\ast}$ and $\Omega_{c}^{\ast} \bar{D}^{\ast}$  are seen in other scattering processes. A similar pattern is found for bound state $\Omega_{c} \bar{D}^{\ast}$ decaying into open channels $\Xi J/\psi$, $\Xi^{\ast} \eta_{c}$, $\Xi^{\ast} J/\psi$, $\Xi_{c}^{\prime} \bar{D}_{s}^{\ast}$ and $\Xi_{c} \bar{D}_{s}^{\ast}$, indicating that bound state $\Omega_{c} \bar{D}^{\ast}$ transitions into a scattering state in the scattering processes of these channels.  For bound states $\Xi_{c}^{\ast} \bar{D}_{s}$ and $\Omega_{c}^{\ast} \bar{D}$, they can decay into open channels $\Xi J/\psi$, $\Xi^{\ast} \eta_{c}$, $\Xi_{c} \bar{D}_{s}^{\ast}$ and  $\Xi J/\psi$, $\Xi^{\ast} \eta_{c}$, $\Xi^{\ast} J/\psi$, $\Xi_{c} \bar{D}_{s}^{\ast}$, respectively. In Fig.~\ref{0.5-1.5-two-coupling-6-9}, bound states $\Xi_{c}^{\ast} \bar{D}_{s}$ and $\Omega_{c}^{\ast} \bar{D}$ both transformed into scattering states in the corresponding open channels scattering phase shifts processes. To further explore the effect of multi-channel coupling on resonance state formation, a six-channel coupling scenario involving five bound states and one open channel is also examined. Fig.~\ref{0.5-1.5-six-coupling} illustrates the absence of resonant behavior in the phase shifts of open channels $\Xi J/\psi$ and $\Xi^{\ast} \eta_{c}$, implying that all five bound states transform into scattering states under six-channel coupling. However, the scattering phase shift of open channel $\Xi_{c} \bar{D}_{s}^{\ast}$  shows pronounced resonance at incident energy around 60 MeV, suggesting that only bound state $\Xi_{c}^{\ast} \bar{D}_{s}$  transforms into a resonant state, while the others transition to scattering states. Table.~\ref{phaseshifts-1.5} provides the masses and decay widths of the three obtained resonance states. The corrected mass and decay width of resonance state $\Xi_{c}^{\ast} \bar{D}_{s}^{\ast}$ are 4749 MeV$-$4755 MeV and 16.4 MeV, respectively. Resonance state $\Omega_{c}^{\ast} \bar{D}^{\ast}$ has a corrected mass of 4771 MeV$-$4772 MeV and a decay width of 18.8 MeV. The corrected mass of resonance state $\Xi_{c}^{\ast} \bar{D}_{s}$ is approximately 4600 MeV, with a decay of around 21.7 MeV. According to the results of Ref.~\cite{Azizi:2021pbh}, a resonance state with a mass around 4600 MeV was identified, consistent with the findings of this study. Nonetheless, the decay width of the resonance determined in this study is somewhat narrower than that reported in Ref.~\cite{Azizi:2021pbh}. In addition, it is also worth comparing these results with the obtained in Ref.~\cite{Ortega:2022uyu} where the resonance states $\Xi_{c}^{\ast} \bar{D}_{s}^{\ast}$ and $\Xi_{c}^{\ast} \bar{D}_{s}$ can be derived from decay channels 
$\Xi^{\ast} \eta_{c}$ and $\Xi J/ \Psi$, $\Xi^{\ast} \eta_{c}$, respectively. Although the masses obtained align within the uncertainty, the decay channels and decay widths of the identified resonances differ, and our results for the width are significantly larger.

%%%%%%%%%%%%%%%%%%%%%%%%%%%%%%%%%%%%%%%%%%%%%%%%%%%%%%%%%%%%%%%%%%%%%%%%%%%%%%%%%%%%%%%%%%%%%%%%%%%%%%%%%%%%%%%%%%%%%%%%

\begin{table}[htb]
    \begin{center}
        \caption{\label{bound-1.5}The binding energies and the masses of every single channel and those of channel coupling for the hidden charm pentaquarks with double strangeness with $J^{P}=3/2^{-}$. The values are provided in units of MeV. }
        \renewcommand\arraystretch{1.5}
        \resizebox{0.48\textwidth}{!} {
        \begin{tabular}{p{1.cm}<\centering p{1.cm}<\centering p{1.cm}<\centering p{1.cm}<\centering p{1.cm}<\centering p{1.cm}<\centering p{1.cm}<\centering p{1.cm}<\centering p{1.cm}<\centering p{1.0cm}<\centering p{1.0cm}<\centering p{1.0cm}<\centering p{1.0cm}<\centering p{1.0cm}<\centering p{1.0cm}<\centering p{1.0cm}<\centering p{1.0cm}<\centering p{1.0cm}<\centering}
            %\begin{tabular}{ccccc|cccc}
            \toprule[1pt]
            
            Channel    & $E_{sc}$  &$E_{th}^{Model}$   &$E_{B}$  &$E_{th}^{Exp}$  &$E_{sc}^{\prime}$  &$E_{cc}/E_{B}$  & $E_{cc}^{\prime}$ \\
            \midrule[1pt]
            $\Xi J/\psi$                              &4364  &4363 &1  &4414  &4415  &\multirow{10}{*}{4364/1}   &\multirow{10}{*}{4415}  \\
            $\Xi^{\ast} \eta_{c}$                     &4482  &4481 &1  &4513  &4514  \\
            $\Xi^{\ast} J/\psi$                       &4485  &4484 &1  &4629  &4630 \\                 
            $\Xi_{c}^{\prime} \bar{D}_{s}^{\ast}$     &4654  &4652 &2  &4689  &4691\\
            $\Xi_{c} \bar{D}_{s}^{\ast}$              &4584  &4583 &1  &4579  &4580\\
            $\Xi_{c}^{\ast} \bar{D}_{s}$              &4649  &4657 &-8 &4613  &4605 \\
            $\Xi_{c}^{\ast} \bar{D}_{s}^{\ast}$       &4637  &4671 &-34&4757  &4723 \\
            $\Omega_{c} \bar{D}^{\ast}$               &4700  &4705 &-5 &4702  &4697\\
            $\Omega_{c}^{\ast} \bar{D}$               &4677  &4682 &-5 &4639  &4634\\
            $\Omega_{c}^{\ast} \bar{D}^{\ast}$        &4709  &4716 &-7 &4777  &4770\\                          
            
            \bottomrule[1pt]
        \end{tabular}
        }
    \end{center}
\end{table}

\begin{table*}[htb]
    \begin{center}
        \caption{\label{phaseshifts-1.5} The masses and decay widths (in the unit of MeV) of resonance states with the difference scattering process with $J^{P}=3/2^{-}$.  }
        \renewcommand\arraystretch{1.5}
        \begin{tabular}{p{2.3cm}<\centering p{1.cm}<\centering p{1.cm}<\centering p{0.5cm}<\centering p{0.01cm}<\centering p{1.cm}<\centering p{1.cm}<\centering p{0.5cm}<\centering p{0.01cm}<\centering p{1.cm}<\centering p{1.cm}<\centering p{0.5cm}<\centering p{0.01cm}<\centering p{1.cm}<\centering p{1.cm}<\centering p{0.5cm}<\centering p{0.2cm}<\centering p{1.cm}<\centering p{1.cm}<\centering p{2.cm}<\centering p{1.cm}<\centering p{1.cm}<\centering p{1.cm}<\centering   p{2.0cm}<\centering  }
            \toprule[1pt]
            \multirow{4}{*}{} & \multicolumn{6}{c}{Two-channel coupling}  &  & \multicolumn{3}{c}{Six-channel coupling} \\
            \cline{2-8} \cline{10-12}
            
            &\multicolumn{3}{c}{$\Xi_{c}^{\ast} \bar{D}_{s}^{\ast}$} &  &\multicolumn{3}{c}{$\Omega_{c}^{\ast}\bar{D}^{\ast}$}  & &\multicolumn{3}{c}{$\Xi_{c}^{\ast} \bar{D}_{s}$}\\
            \cline{2-4}\cline{6-8}\cline{10-12}
            Open channels & $M_{R}^{th}$ & $M_{R}$ & $\Gamma_{i}$  &     &$M_{R}^{th}$ & $M_{R}$ &$\Gamma_{i}$  &     &$M_{R}^{th}$ & $M_{R}$ &$\Gamma_{i}$     \\
            \midrule[1pt]
            $\Xi J/\psi$           &4668  &4754     &10.2  &   &\ldots   &\ldots &\ldots   & &\ldots   &\ldots &\ldots\\
            $\Xi^{\ast} \eta_{c}$  &\ldots&\ldots   &\ldots&   &\ldots   &\ldots &\ldots   & &\ldots   &\ldots &\ldots\\
            $\Xi^{\ast} J/\psi$    &4663  &4749     &4.5   &   &4711     &4772   &15.6     & &\ldots   &\ldots &\ldots\\
            $\Xi_{c}^{\prime} \bar{D}_{s}^{\ast}$&\ldots&\ldots&\ldots &      &4710  &4771 &3.2 &  &\ldots   &\ldots &\ldots\\
            $\Xi_{c} \bar{D}_{s}^{\ast}$ &4669 &4755   &1.7&   &\ldots &\ldots&\ldots &  &4644   &4600 &21.7  \\
            $\Gamma_{Total}$             &     &       &16.4   &   &   &   &18.8  &  &  &  &21.7\\
            \bottomrule[1pt]
        \end{tabular}
    \end{center}
\end{table*}
%%%%%%%%%%%%%%%%%%%%%%%%%%%%%%%%%%%%%%%%%%%%%%%%%%%%%%%%%%%%%%%%%%%%%%%%%%%%%%%%%%%%%%%%%%%%%%%%%%%%%%%%%%%%%%%%

\subsection{the $J^{P}=\frac{5}{2}^{-}$}
\begin{figure}[t]
  \includegraphics[scale=0.30]{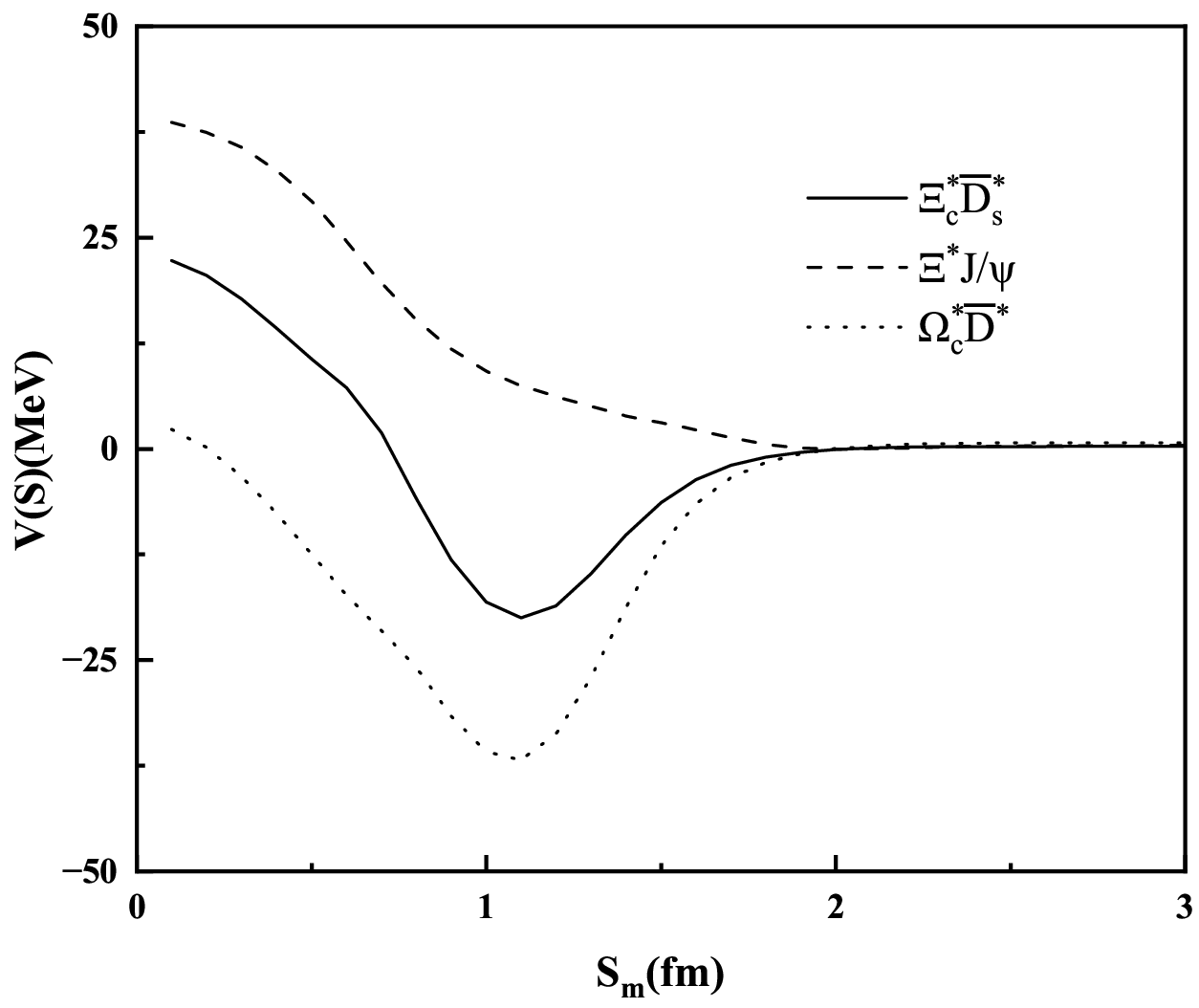}
    \caption{ The effective potentials defined in Eq.~(\ref{Eq:PotentialV}) for different channels of the hidden charm and double strangeness pentaquark systems with $J^{P}=5/2^{-}$ in QDCSM. }
    \label{Veff-2.5}
\end{figure}

%%%%%%%%%%%%%%%%%%%%%%%%%%%%%%%%%%%%%%%%%%%%%%%%%%%%%%%%%%%%%%%%%%%%%%%%%%%%%%%%%%%%%%%%%%%%%%%%%%%%%%%%%%%%%%%%%
For the $J^{P}=5/2^{-}$, it is evident from Table~\ref{channels} that there are three physical channels: $\Xi^{\ast} J/\psi$, $\Xi_{c}^{\ast} \bar{D}_{s}^{\ast}$ and $\Omega_{c}^{\ast} \bar{D}^{\ast}$. Fig.~\ref{Veff-2.5} presents the variation trend of the effective potentials for these physical channels. As can be seen from Fig.~\ref{Veff-2.5}, the effective potential of  $\Xi^{\ast} J/\psi$ exhibits a repulsive nature, whereas $\Xi_{c}^{\ast} \bar{D}_{s}^{\ast}$ and $\Omega_{c}^{\ast} \bar{D}^{\ast}$ display attractive characteristics. In comparing the two channels with attractive potentials, the effective attractive strength of $\Omega_{c}^{\ast}\bar{D}^{\ast}$ is significantly stronger than that of $\Xi^{\ast} J/\psi$. This feature suggests that  $\Omega_{c}^{\ast} \bar{D}^{\ast}$ is more probable to form a bound state.

Based on the analysis of the effective potentials of each physical channel, Table.~\ref{bound-2.5} summarizes the estimated results of dynamical bound states under single-channel and multi-channel coupling. Due to its effective repulsive characteristics, the single-channel estimate for $\Xi^{\ast} J/\psi$ exceeds its theoretical threshold. The dynamical bound state estimate for $\Xi_{c}^{\ast} \bar{D}_{s}^{\ast}$  is similar to that of $\Xi^{\ast} J/\psi$, however,  $\Xi_{c}^{\ast} \bar{D}_{s}^{\ast}$  fails to form a bound state due to its relatively weak effective attraction. $\Omega_{c}^{\ast} \bar{D}^{\ast}$ forms a bound state with a binding energy of -7 MeV, resulting from the strong attractive interactions between hadrons. With further channel coupling considerations, the estimated result is about 4630 MeV, which is higher than the threshold of the lowest $\Omega_{c}^{\ast} \bar{D}^{\ast}$, suggesting that no bound state is derived from the hidden-charm and double strangeness pentaquark system with $J^{P}=5/2^{-}$.

\begin{figure}[htp]
    \includegraphics[scale=0.35]{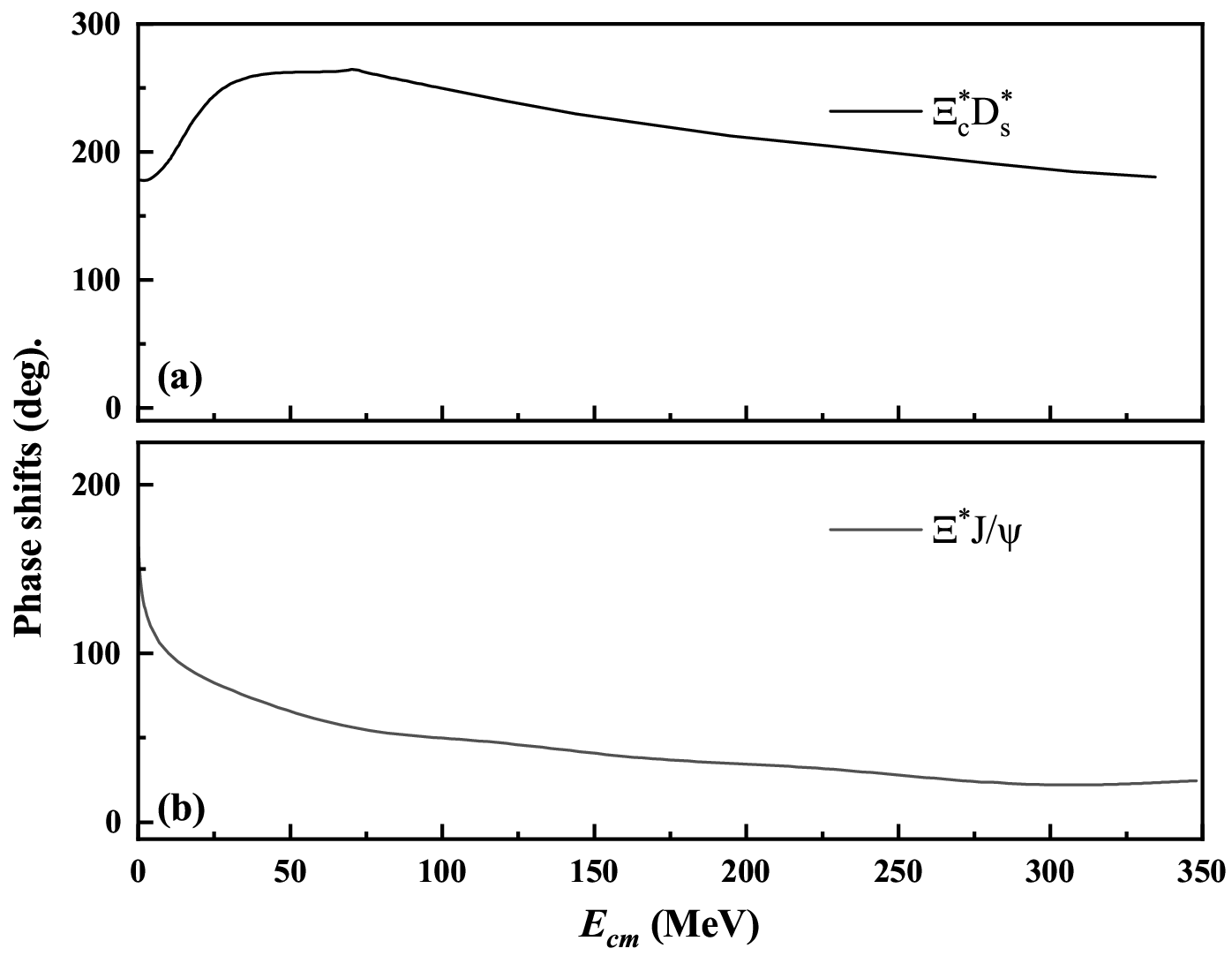}
    \caption{The phase shifts of the open channels with two-channel coupling for $J^{P}=5/2^{-}$ in QDCSM; (a)denotes two-channel coupling with $\Omega_{c}^{\ast}\bar{D}^{\ast}$ and $\Xi_{c}^{\ast}\bar{D}_{s}^{\ast}$, (b) stands for two-channel coupling with $\Omega_{c}^{\ast}\bar{D}^{\ast}$ and $\Xi^{\ast} J/\psi$.}
    \label{0.5-2.5-two-coupling}
\end{figure}

The dynamical bound state estimation indicates the existence of a bound state, denoted as $\Omega_{c}^{\ast} \bar{D}^{\ast}$, within the hidden-charm and double strangeness pentaquark system with $J^{P}=5/2^{-}$. As only the S-wave scenario is analyzed here, $\Omega_{c}^{\ast} \bar{D}^{\ast}$ can decay exclusively into open channels $\Xi^{\ast} J/\psi$ and $\Xi_{c}^{\ast} \bar{D}_{s}^{\ast}$. To assess if $\Omega_{c}^{\ast} \bar{D}^{\ast}$ forms a resonance state within the scattering phase shifts across different open channels, a two-channel coupling between $\Omega_{c}^{\ast} \bar{D}^{\ast}$ and each open channel is explored. Fig.~\ref{0.5-2.5-two-coupling} depicts the scattering phase behavior of $\Omega_{c}^{\ast} \bar{D}^{\ast}$ in relation to open channels $\Xi^{\ast} J/\psi$ and $\Xi_{c}^{\ast} \bar{D}_{s}^{\ast}$. As shown in Fig.~\ref{0.5-2.5-two-coupling}, no significant resonance signals are observed in either scattering process of the open channel, revealing that $\Omega_{c}^{\ast} \bar{D}^{\ast}$ transitions to the scattering state under these conditions.

%%%%%%%%%%%%%%%%%%%%%%%%%%%%%%%%%%%%%%%%%%%%%%%%%%%%%%%%%%%%%%%%%%%%%%%%%%%%%%%%%%%%%%%%%%%%%%%%%%%%%%%%%%%%%%%%%%%%%%

\begin{table}[htb]
    \begin{center}
        \caption{\label{bound-2.5}The binding energies and the masses of every single channel and those of channel coupling for the hidden charm pentaquarks with double strangeness with $J^{P}=5/2^{-}$. The values are provided in units of MeV. }
        \renewcommand\arraystretch{1.5}
        \resizebox{0.48\textwidth}{!} {
        \begin{tabular}{p{1.cm}<\centering p{1.cm}<\centering p{1.cm}<\centering p{1.cm}<\centering p{1.cm}<\centering p{1.cm}<\centering p{1.cm}<\centering p{1.cm}<\centering p{1.cm}<\centering p{1.0cm}<\centering p{1.0cm}<\centering p{1.0cm}<\centering p{1.0cm}<\centering p{1.0cm}<\centering p{1.0cm}<\centering p{1.0cm}<\centering p{1.0cm}<\centering p{1.0cm}<\centering}
            %\begin{tabular}{ccccc|cccc}
            \toprule[1pt]
            
            Channel    & $E_{sc}$  &$E_{th}^{Model}$   &$E_{B}$  &$E_{th}^{Exp}$  &$E_{sc}^{\prime}$  &$E_{cc}/E_{B}$  & $E_{cc}^{\prime}$ \\
            \midrule[1pt]
            $\Xi^{\ast} J/\psi$                       &4485  &4484 & 1 &4629  &4630  &\multirow{3}{*}{4485/1}   &\multirow{3}{*}{4630} \\     
            $\Xi_{c}^{\ast} \bar{D}_{s}^{\ast}$       &4673  &4671 & 2 &4757  &4759 \\
            $\Omega_{c}^{\ast} \bar{D}^{\ast}$        &4709  &4716 &-7 &4777  &4770 \\                          
            \bottomrule[1pt]
        \end{tabular}
        }
    \end{center}
\end{table}

\begin{table}[htb]
    \begin{center}
        \caption{\label{summarize} All obtained resonance states and possible decay channels under S-wave approximation in the hidden-charm and doubly strangeness system. The values are provided in units of MeV.  }
        \renewcommand\arraystretch{1.5}
        \resizebox{0.48\textwidth}{!} {
        \begin{tabular}{p{0.6cm}<\centering p{2.3cm}<\centering p{2.6cm}<\centering p{1.5cm}<\centering p{1.5cm}<\centering p{0.5cm}<\centering p{0.5cm}<\centering p{0.5cm}<\centering p{1.5cm}<\centering p{1.5cm}<\centering p{1.cm}<\centering p{0.5cm}<\centering p{0.01cm}<\centering p{1.cm}<\centering p{1.cm}<\centering p{0.5cm}<\centering p{0.2cm}<\centering p{1.cm}<\centering p{1.cm}<\centering p{2.cm}<\centering p{1.cm}<\centering p{1.cm}<\centering p{1.cm}<\centering   p{2.0cm}<\centering  }
            \toprule[1pt]
            $J^{P}$ & Resonance state   & Decay channels    &Mass      & $\Gamma_{Total}$ \\
            \midrule[1pt]
            \multirow{2}{*}{$\frac{1}{2}^{-}$} &$\Xi^{\prime}_{c}\bar{D}_{s}^{\ast}$   &$\Xi \eta_{c}$, $\Xi J/\psi$, $\Xi^{\ast} J/\psi$, $\Xi_{c}\bar{D}_{s}$, $\Xi_{c}\bar{D}_{s}^{\ast}$ &4682-4688 &6.4-24.2\\
            &$\Xi^{\ast}_{c}\bar{D}_{s}^{\ast}$   &$\Xi \eta_{c}$,$\Xi J/\psi$, $\Xi_{c}\bar{D}_{s}$ &4751-4756 &3.2-18.7\\
             \midrule[1pt]
            \multirow{3}{*}{$\frac{3}{2}^{-}$} &$\Xi^{\ast}_{c}\bar{D}_{s}$   & $\Xi_{c}\bar{D}_{s}^{\ast}$   &4600 &21.7\\
            &$\Xi^{\ast}_{c}\bar{D}_{s}^{\ast}$   &$\Xi J/\psi$, $\Xi^{\ast} J/\psi$, $\Xi_{c}\bar{D}_{s}^{\ast}$   &4749-4755 &16.4\\
            &$\Omega_{c}^{\ast} \bar{D}^{\ast}$   &$\Xi^{\ast} J/\psi$, $\Xi_{c}^{\prime}\bar{D}_{s}^{\ast}$ &4771-4772 &18.8\\
            \bottomrule[1pt]
        \end{tabular}
        }
    \end{center}
\end{table}

%%%%%%%%%%%%%%%%%%%%%%%%%%%%%%%%%%%%%%%%%%%%%%%%%%%%%%%%%%%%%%%%%%%%%%%%%%%%%%%%%%%%%%%%%%%%%%%%%%%%%%%%%%%%%%
\section{Summary\label{sum}}
In this study, the properties of hidden-charm double-strange pentaquark systems with various quantum numbers are investigated systematically by utilizing the RGM within the QDCSM framework. To assess the possible existence of bound or resonance states in the hidden-charm double-strange pentaquark systems, the effective potentials between hadrons for different quantum numbers are estimated to gain a deeper insight into their interaction characteristics. Besides, given the dynamical bound state estimation, the likelihood of bound state formation in hidden-charm double-strange pentaquark systems is evaluated through single-channel and multi-channel coupling estimations. The single-channel estimation results suggest that two bound states, $\Xi^{\prime}_{c}\bar{D}_{s}^{\ast}$ with an estimated mass of 4681 MeV and $\Xi^{\ast}_{c}\bar{D}_{s}^{\ast}$ with an estimated mass of 4744 MeV, are identified with $J^{P}=1/2^{-}$; for $J^{P}=3/2^{-}$, five bound states, labeled $\Xi^{\ast}_{c}\bar{D}_{s}$, $\Xi^{\ast}_{c}\bar{D}_{s}^{\ast}$, $\Omega_{c}\bar{D}^{\ast}$, $\Omega_{c}^{\ast}\bar{D}$ and $\Omega^{\ast}_{c}\bar{D}^{\ast}$, are spotted with estimated masses of 4605 MeV, 4723 MeV, 4697 MeV, 4634 MeV, and 4770 MeV, respectively. However, after accounting for the effects of multi-channel coupling, no bound states are detected for any quantum number.

Based on the above-bound state estimation results, a detailed examination of scattering phase shifts in open channels is required to confirm the existence of possible resonance states. Table.~\ref{summarize} summarizes the relevant estimation results, showing five observed resonance states for $J^{P}=1/2^{-}$ and $J^{P}=3/2^{-}$. For the $J^{P}=1/2^{-}$,  the resonance state $\Xi^{\prime}_{c}\bar{D}_{s}^{\ast}$ has a mass of 4682 MeV-4688 MeV and a decay width of 6.4 MeV$-$24.2 MeV, observed in scattering phase shifts of $\Xi \eta_{c}$, $\Xi J/\psi$, $\Xi^{\ast} J/\psi$, $\Xi_{c}\bar{D}_{s}$, and $\Xi_{c}\bar{D}_{s}^{\ast}$; similarly, the resonance state $\Xi^{\ast}_{c}\bar{D}_{s}^{\ast}$ with mass and decay width of about 4751 MeV$-$4756 MeV and 3.2 MeV$-$18.7 MeV appears in the scattering processes of open channels $\Xi \eta_{c}$, $\Xi J/\psi$, and $\Xi_{c}\bar{D}_{s}$. In the sector with $J^{P}=3/2^{-}$, three resonance states, $\Xi^{\ast}_{c}\bar{D}_{s}$ with a mass of 4600 MeV and width of 21.7 MeV in the scattering process of open channel $\Xi_{c}\bar{D}_{s}^{\ast}$, $\Xi^{\ast}_{c}\bar{D}_{s}^{\ast}$  with a mass of 4749 MeV$-$4755 MeV and width of 16.4 MeV in the scattering process of open channels $\Xi J/\psi$, $\Xi^{\ast} J/\psi$, and  $\Xi_{c}\bar{D}_{s}^{\ast}$, and $\Omega_{c}^{\ast} \bar{D}^{\ast}$ with a mass of 4771 MeV$-$4772 MeV and width of 18.8 MeV in the scattering process of open channels  $\Xi^{\ast} J/\psi$ and $\Xi_{c}^{\prime}\bar{D}_{s}^{\ast}$, are detected.  It is worth noting that the current analysis of hidden-charm double-strange pentaquark systems with different quantum numbers is restricted to the S-wave condition. A more comprehensive investigation could introduce higher partial-wave coupling through tensor-force interactions, which will be addressed in future studies. Finally, we hope that the systematic analysis of the hidden-charm double-strange pentaquark system provides a reliable theoretical foundation for experimental exploration.

\acknowledgments{This work is supported partly by the National Natural Science Foundation of China under
    Contract No. 12175037, No. 12335001, No. 11775118 and No. 11535005, and is also supported by the Natural Science Foundation for Youths of Henan Province  No. 252300421781.  School-Level Research Projects of Henan Normal unversity (No. 20240304) also supported this work. Besides, Jiangsu Provincial Natural Science Foundation Project (No. BK20221166), National Youth Fund: No. 12205125 and School-Level Research Projects of Yancheng Institute of Technology (No. xjr2022039) also supported this work.}

\bibliography{uss-ccbar}

\end{document}